\documentclass[aip,amsmath,amssymb,reprint]{revtex4-1}
\usepackage{amsthm}
\usepackage[english]{babel}
\usepackage[utf8]{inputenc}
\usepackage[colorinlistoftodos, color=green!40, prependcaption]{todonotes}
\usepackage{mathtools}
\usepackage{physics}
\usepackage{xcolor}
\usepackage{graphicx}
\usepackage[left=23mm,right=13mm,top=35mm,columnsep=15pt]{geometry} 
\usepackage{adjustbox}
\usepackage{placeins}
\usepackage[T1]{fontenc}
\usepackage{lipsum}
\usepackage{csquotes}
\usepackage{tikz}
\usepackage{bm}
\usepackage{braket}
\usepackage{multirow}
\usepackage{qcircuit}
\usepackage{siunitx}
\usepackage{ulem}

\usepackage{dcolumn}
\usepackage{mathptmx}
\usepackage{etoolbox}

\usepackage{hyperref}
\hypersetup{
    colorlinks=true,
    linkcolor=blue,
    urlcolor=blue,
    citecolor=blue
    }
\usepackage[capitalise]{cleveref}

\newcommand{\R}{Ref.~}

\newcommand{\fig}{Fig.~}

\makeatletter
\def\@email#1#2{%
 \endgroup
 \patchcmd{\titleblock@produce}
  {\frontmatter@RRAPformat}
  {\frontmatter@RRAPformat{\produce@RRAP{*#1\href{mailto:#2}{#2}}}\frontmatter@RRAPformat}
  {}{}
}%
\makeatother

\begin{document}

\title{Efficient Generation of Multi-partite Entanglement between Non-local Superconducting Qubits using Classical Feedback}

\author{Akel Hashim}
    \email{ahashim@berkeley.edu.}
    \thanks{These authors contributed equally to this work.}
    \affiliation{Quantum Nanoelectronics Laboratory, Department of Physics, University of California at Berkeley, Berkeley, CA 94720, USA}
\author{Ming Yuan}
    \thanks{These authors contributed equally to this work.}
    \affiliation{Pritzker School of Molecular Engineering, University of Chicago, Chicago, IL 60637, USA}
\author{Pranav Gokhale}
    \affiliation{Infleqtion, Chicago, IL 60604, USA}
\author{Larry Chen}
    \affiliation{Quantum Nanoelectronics Laboratory, Department of Physics, University of California at Berkeley, Berkeley, CA 94720, USA}
\author{Christian Jünger}
    \thanks{Now at YQuantum, Parkstrasse 1, CH-5234 Villigen, Switzerland.}
    \affiliation{Quantum Nanoelectronics Laboratory, Department of Physics, University of California at Berkeley, Berkeley, CA 94720, USA}
    \affiliation{Applied Math and Computational Research Division, Lawrence Berkeley National Lab, Berkeley, CA 94720, USA}
\author{Neelay Fruitwala}
\author{Yilun Xu}
\author{Gang Huang}
    \affiliation{Accelerator Technology and Applied Physics Division, Lawrence Berkeley National Lab, Berkeley, CA 94720, USA}
\author{Kasra Nowrouzi}
    \affiliation{Applied Math and Computational Research Division, Lawrence Berkeley National Lab, Berkeley, CA 94720, USA}
\author{Liang Jiang}
    \affiliation{Pritzker School of Molecular Engineering, University of Chicago, Chicago, IL 60637, USA}
\author{Irfan Siddiqi}
    \affiliation{Quantum Nanoelectronics Laboratory, Department of Physics, University of California at Berkeley, Berkeley, CA 94720, USA}
    \affiliation{Applied Math and Computational Research Division, Lawrence Berkeley National Lab, Berkeley, CA 94720, USA}
    \affiliation{Materials Sciences Division, Lawrence Berkeley National Lab, Berkeley, CA 94720, USA}
    
\date{\today}

\begin{abstract}
Quantum entanglement is one of the primary features which distinguishes quantum computers from classical computers. In gate-based quantum computing, the creation of entangled states or the distribution of entanglement across a quantum processor often requires circuit depths which grow with the number of entangled qubits. However, in teleportation-based quantum computing, one can deterministically generate entangled states with a circuit depth that is constant in the number of qubits, provided that one has access to an entangled resource state, the ability to perform mid-circuit measurements, and can rapidly transmit classical information. In this work, aided by fast classical field programmable gate array-based control hardware with a feedback latency of only 150 ns, we explore the utility of teleportation-based protocols for generating non-local, multi-partite entanglement between superconducting qubits. First, we demonstrate well-known protocols for generating Greenberger-Horne-Zeilinger (GHZ) states and non-local CNOT gates in constant depth. Next, we utilize both protocols for implementing a quantum fan-out gate in constant depth among three non-local qubits (i.e., controlled-NOT-NOT). Finally, we demonstrate deterministic state teleportation and entanglement swapping between qubits on opposite sides of our quantum processor. Throughout this work, we find that the fidelity of our teleportation-based protocols is limited by measurement-induced dephasing on idling spectator qubits. Therefore, our work serves as a useful study of the current benefits and limitations of teleportation-based protocols on contemporary superconducting quantum processors.
\end{abstract}


\maketitle

\section{Introduction}\label{sec:intro}

Quantum entanglement plays a central role in nearly all quantum applications. In gate-based quantum computing, entanglement is generated by multi-qubit unitary operations (i.e., gates), which can be used to prepare multi-partite entangled states. In systems with all-to-all connectivity, generating multi-partite entanglement can be trivially accomplished by means of two-qubit gates \cite{moses2023race}. However, in systems with limited connectivity (e.g., superconducting circuits), generating non-local multi-partite entanglement by unitary operators can be costly due to the need for SWAP networks for swapping states between neighboring qubits \cite{o2019generalized, hashim2022optimized}. However, long-range entangled states can be efficiently prepared using quantum teleportation-based protocols \cite{bennett1993teleporting, gottesman1999demonstrating} in a hardware-agnostic manner. Quantum teleportation is a method for transmitting a quantum state from one qubit to another, provided that they each share half of a Bell state and can transmit classical information. When embedded within a larger (unitary) quantum circuit for performing adaptive control of entangled states, teleportation protocols necessitate the use of mid-circuit measurements (MCMs) and fast classical feed-forward operations conditioned on the results of the measured outcomes. Using adaptive circuits, it has been shown that no purely unitary circuit can prepare the same entangled state for the same circuit depth and connectivity \cite{foss2023experimental}. 
For example, a variety of novel entangled states can be deterministically prepared with adaptive circuits in constant depth, including certain classes of matrix product states\cite{smith_constant-depth_2024}, non-Abelian or other exotic topologically ordered states\cite{bravyi2022adaptive, tantivasadakarn2022hierarchy, tantivasadakarn2022shortest, lu2022measurement, lu2023mixed,verresen_efficiently_2022}, and W or low-weight Dicke states\cite{buhrman_state_2024-1, niu2024ac} as well.
Moreover, adaptive circuits also provide a significant sampling overhead reduction in certifying multi-partite quantum states\cite{gupta_few_2025}.
Finally, as the size of quantum computers continues to grow, adopting modular computing architectures \cite{monroe2014large} will be necessary. In such architectures, quantum teleportation protocols \cite{gottesman1999demonstrating, bennett1993teleporting, jiang2007distributed} will be useful tools for transmitting quantum information from one system to another.

In the noisy intermediate-scale quantum (NISQ) era \cite{preskill2018quantum}, qubit coherence times and gate fidelities limit the circuit depth with which we can perform useful computations. For superconducting qubits, if long-range entanglement is needed between qubits which are not physically connected, swapping information requires $\mathcal{O}(n)$ circuit depth for $n$ qubits \cite{hashim2022optimized}. Therefore, generating multi-partite entanglement between non-local qubits becomes intractable if the available size of contemporary systems grows faster than the fidelity of multi-qubit gates can be improved. However, 
as mentioned above,
non-local, multi-partite entangled states can be prepared in constant depth using teleportation-based protocols, as long as one has access to mid-circuit measurements and adaptive feedback. These entangled states can further be used as a resource for other teleportation-based protocols (e.g., generating non-local entangling gates). 

In this work, we utilize open-source control hardware based on field programmable gate arrays (FPGAs) with low feedback latency ($\sim$150 ns) \cite{xu2021qubic, xu2023qubic, fruitwala2024distributed} for exploring the utility of teleportation-based protocols on an eight-qubit superconducting quantum processor (see Appendix). We first demonstrate well-known procedures for preparing Greenberger–Horne–Zeilinger (GHZ) states and implementing non-local teleportation-based controlled-NOT (CNOT) gates in constant depth. Then, we show how one can utilize both protocols to implement an unbounded fan-out gate in constant depth, which we use to create a controlled-NOT-NOT (CNOT-NOT) gate between three unconnected qubits. Finally, we show how one can deterministically teleport quantum states or generate Bell pairs between distant qubits in constant-depth, and demonstrate these protocols between qubits on opposite sides of our quantum processor. Throughout this work, we explore the extent to which MCMs dephase idling spectator qubits, and find that the measurement-induced dephasing \cite{gambetta2006qubit} strongly depends on the spectral resolution of the readout resonators coupled to each qubit, which limits the size and complexity of the protocols that we can implement on our current processor. This work serves as a useful study for understanding how teleportation-based protocols can be utilized as a resource within larger (unitary-based) quantum circuits, and what limitations need to be overcome for utilizing such protocols for superconducting qubits at scale.
We believe that adaptive circuits will become a useful tool for state-preparation and non-local gates when the circuit depths of the equivalent unitary circuits are prohibitive on systems with limited connectivity. 

\section{GHZ State Preparation}\label{sec:ghz}

\begin{figure*}[t]
    \centering
    \includegraphics[width=2\columnwidth]{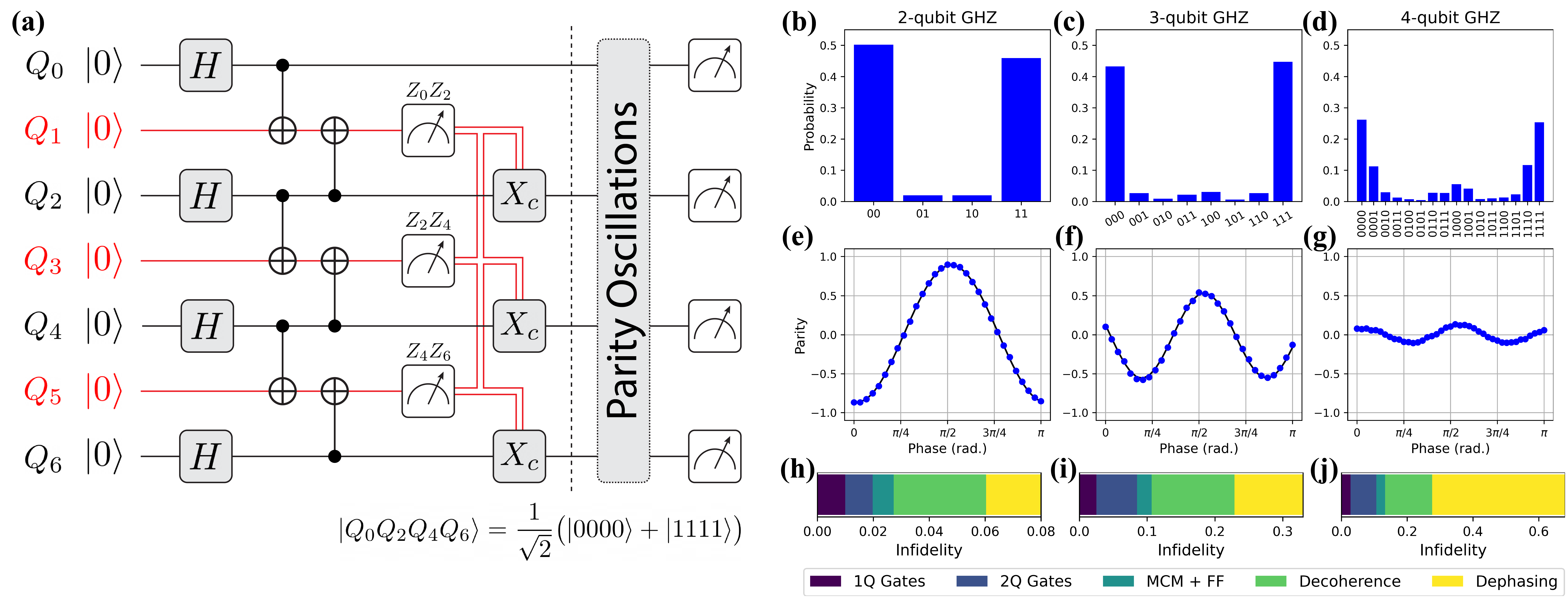}
    \caption{\textbf{GHZ State Preparation in Constant Depth.}
    \textbf{(a)} Adaptive circuit for preparing a four-qubit GHZ state on non-local qubits. Data qubits (black) are prepared in a superposition state with a Hadamard (H) gate and entangled with ancillae qubits (red) in a pair-wise fashion. Mid-circuit measurements of the ancillae qubits can be used to determine the parity of the data qubits, which can be decoded to determine which data qubits should be flipped via conditional $X$ gates ($X_c$) to prepare all four in a GHZ state. 
    Measurements of the GHZ state in the computational basis for \textbf{(b)} two qubits, \textbf{(c)} three qubits, and \textbf{(d)} four qubits.
    Parity oscillations of the GHZ state for \textbf{(e)} two qubits, \textbf{(f)} three qubits, and \textbf{(g)} four qubits. From these results we calculate a GHZ state prep fidelity of $F_{\ket{\text{GHZ}_2}} = 0.92(1)$, $F_{\ket{\text{GHZ}_3}} = 0.67(2)$, and $F_{\ket{\text{GHZ}_4}} = 0.32(3)$. 
    Error budgets for the \textbf{(h)} two-, \textbf{(i)} three-, and \textbf{(j)} four-qubit GHZ state preparation experiments. Dephasing on the spectator qubits becomes the dominant source of error as the number of qubits is increased.
    }
    \label{fig:ghz}
\end{figure*}

The first task we demonstrate is the preparation of a GHZ state between non-nearest neighbor qubits in constant depth. GHZ states are highly entangled states that play an important role in quantum foundations \cite{greenberger1989going, mermin1990quantum, caves2002unknown}, quantum metrology \cite{eldredge2018optimal}, and fault-tolerant quantum computation \cite{Shor1996}. To prepare an $n$-qubit GHZ state with one- and two-qubit gates requires at least $\mathcal{O}(\log n)$ circuit depth even with all-to-all connectivity \cite{watts_exponential_2019}. However, if we are allowed to perform MCMs and implement gate operations on the unmeasured qubits based on the classical outcomes, the depth can be reduced to constant regardless of the size of the GHZ state. The idea for constant depth GHZ state preparation can be easily understood via the stabilizer formalism. The GHZ state $\ket{\mathrm{GHZ}}=\frac{1}{\sqrt{2}}(\ket{0}^{\otimes n} + \ket{1}^{\otimes n})$ is the eigenstate for a set of stabilizer generators $\langle X_1X_2\dots X_n, Z_1Z_2, \dots, Z_j Z_{j+1}, \dots, Z_{n-1}Z_n\rangle$ with eigenvalue $+1$. Therefore, one can initialize all data qubits in $\ket{+}^{\otimes n}$ state and use ancillae qubits in the middle of two data qubits to measure $ZZ$ type stabilizers in parallel. If all the measurement outcomes are $+1$, then the state among those data qubits should be in the standard form of the GHZ state, provided that there is no readout error. On the other hand, if some of the measurement outcomes are $-1$, this indicates that there exist some domain walls \footnote{The term domain wall here indicates the location between two sequences of qubits with different parities (qubits in each of the two sequences have the same parity)} between adjacent data qubits and therefore we need an extra layer of single-qubit $X$ gates acting on certain qubits to recover the standard GHZ state. While this protocol has already been demonstrated in trapped ions \cite{moses2023race}, cold atoms \cite{bluvstein2024logical}, and superconducting qubits \cite{baumer2023efficient}, we similarly show it here as it not only serves as the backbone for other tasks like the constant-depth fan-out gate implementation, but also provides one figure of merit to benchmark the performance of measurement- and teleportation-based protocols in our platform.

The circuit schematic for preparing a four-qubit GHZ state in constant depth on our eight-qubit processor is shown in \fig\ref{fig:ghz}(a). Here, four data qubits (black) are entangled with three ancillae qubits (red) in an alternating fashion. By performing an MCM of the ancillae qubits, we can learn about the parity of each pair of data qubits, and the results of the parity measurements can be decoded in real-time to determine which data qubit(s) experienced a bit-flip (see Appendix \ref{sec:LUTs}). In Figs.~\ref{fig:ghz}(b)--(d), we plot the raw (marginalized) outputs of measurements made in the computational basis for two-, three-, and four-qubit GHZ states prepared among data qubits. While the classical outputs resemble GHZ states (albeit very noisy in the four-qubit case), we use parity oscillations \cite{sackett2000experimental, leibfried2005creation, monz201114} to verify whether or not we have generated genuine entanglement among our data qubits. The fidelity of an $n$-qubit GHZ state state can subsequently be computed via
\begin{equation}\label{eq:fid_po}
    F_{\ket{\text{GHZ}_n}} = \frac{1}{2} \left( P_{\ket{0}^{\otimes n}} + P_{\ket{1}^{\otimes n}} + C \right) ~,
\end{equation}
where $P_{\ket{0}^{\otimes n}}$ is the population of the $\ket{0}^{\otimes n}$ state, $P_{\ket{1}^{\otimes n}}$ is the population of the $\ket{1}^{\otimes n}$ state, and $C$ is the amplitude of oscillations of the measured parity (i.e., the coherence of the state) \footnote{Equation \ref{eq:fid_po} is quantitatively similar to the state fidelity of a GHZ state, such as what would be estimated using quantum state tomography \cite{hashim2024practical}.}. If $F_{\ket{\text{GHZ}_n}} > 0.5$, this represents a case of genuine entanglement between $n$ qubits \cite{leibfried2005creation, dur2001multiparticle, guhne2010separability}. For our constant-depth creation of GHZ states, we measure a fidelity of $F_{\ket{\text{GHZ}_2}} = 0.92(1)$, $F_{\ket{\text{GHZ}_3}} = 0.67(2)$, and $F_{\ket{\text{GHZ}_4}} = 0.32(3)$ for the two-, three-, and four-qubit GHZ states, respectively. Therefore, while we observe genuine entanglement for the two- and three-qubit GHZ states, the four-qubit GHZ fails the above criteria.

In general, the fidelity of states prepared via teleportation- and measurement-based protocols can be limited by many sources of errors.
In Figs.~\ref{fig:ghz}(h)--(j), we plot error budgets for the two-, three-, and four-qubit GHZ state preparation experiments constructed using a back-of-the-envelope calculation from the various error rates and decoherence times listed in Appendix \ref{sec:appendix}. We broadly divide the error budgets into gate errors from single-qubit gates (``1Q Gates''), gate errors from two-qubit gates (``2Q Gates''), errors due to mid-circuit measurements and feed-forward operations (``MCM + FF''), decoherence due to $T_1$ decay, and dephasing\footnote{We perform readout correction on the classical output distributions, so we do not include errors due to terminating measurements in our error budgets}. For example, we observe that as we increase the number of qubits in the experiment, dephasing errors grow (both in absolute value and relative to the rest of the error sources) to dominate the error budget.
In our case, the dephasing errors are likely due to measurement-induced dephasing on spectator qubits (see Appendix \ref{sec:meas_dephasing}). Because superconducting qubits are measured via dispersive coupling to a readout resonator, depending on the frequency separation of the resonators of neighboring qubits and the static coupling between nearest neighbors, the measurement of an ancilla qubit can result in the dephasing of a spectator qubit. If the dephasing is weak, it can be mitigated via dynamical decoupling. However, if the readout resonator of the ancilla qubit is close in frequency to the readout resonator of the data qubit, then measurement of the ancilla qubit can indirectly measure the data qubit, in which case phase coherence of the data qubit is completely lost after the MCM (we discuss this in detail in Appendix \ref{sec:meas_dephasing}). This dephasing can be mitigated by ensuring that ancilla qubits are always those whose resonators are spectrally well resolved from the resonators of the data qubits, but this can sometimes necessitate SWAP gates between qubits, which increases the contribution from gate errors in the preparation circuit.

As noted above, the task of preparing GHZ states in constant depth using adaptive circuits has already been demonstrated in a number of different hardware platforms \cite{moses2023race, bluvstein2024logical, baumer2023efficient}. In superconducting qubits, \R\cite{baumer2023efficient} observes genuine entanglement in up to 8 qubits using adaptive circuits, genuine entanglement in up 13 qubits using just MCMs and updating the Pauli frame in post-processing, and genuine entanglement in up to 17 qubits using unitary circuits; all three versions fall short of state-of-the-art fidelities for GHZ states in superconducting qubits \cite{mooney2021generation}. They highlight that their adaptive-based protocols were mainly limited by idling errors during MCM and feed-forward, compared to gate errors in unitary circuits. Thus, while they were able to successfully deploy the protocol on more qubits than we have demonstrated above, they did not show any improvement over unitary-based GHZ state preparation, limiting the utility of the protocol on contemporary superconducting quantum processors. In trapped ions, \R\cite{moses2023race} successfully prepares a 32-qubit GHZ state using adaptive circuits with a fidelity of 0.74(1), but found that this did not outperform the unitary-based GHZ state [which had a fidelity of 0.82(1)] due to the fact that the adaptive circuit actually required more two-qubit gates and measurements than the equivalent unitary circuit; thus, in systems with all-to-all connectivity, adaptive circuits may not provide any benefit over unitary circuits. Finally, in cold atoms, \R\cite{bluvstein2024logical} reports creating a three-qubit logical GHZ state using adaptive circuits with a fidelity of 0.77(2). They note that this fidelity is mainly limited by the underlying GHZ state preparations of the physical qubits, since atomic-based systems require shuttling their atoms between entangling zones, storage zones, and readout zones.

\section{Teleportation-based CNOT}\label{sec:cnot}

\begin{figure}[!th]
    \centering
    \includegraphics[width=\columnwidth]{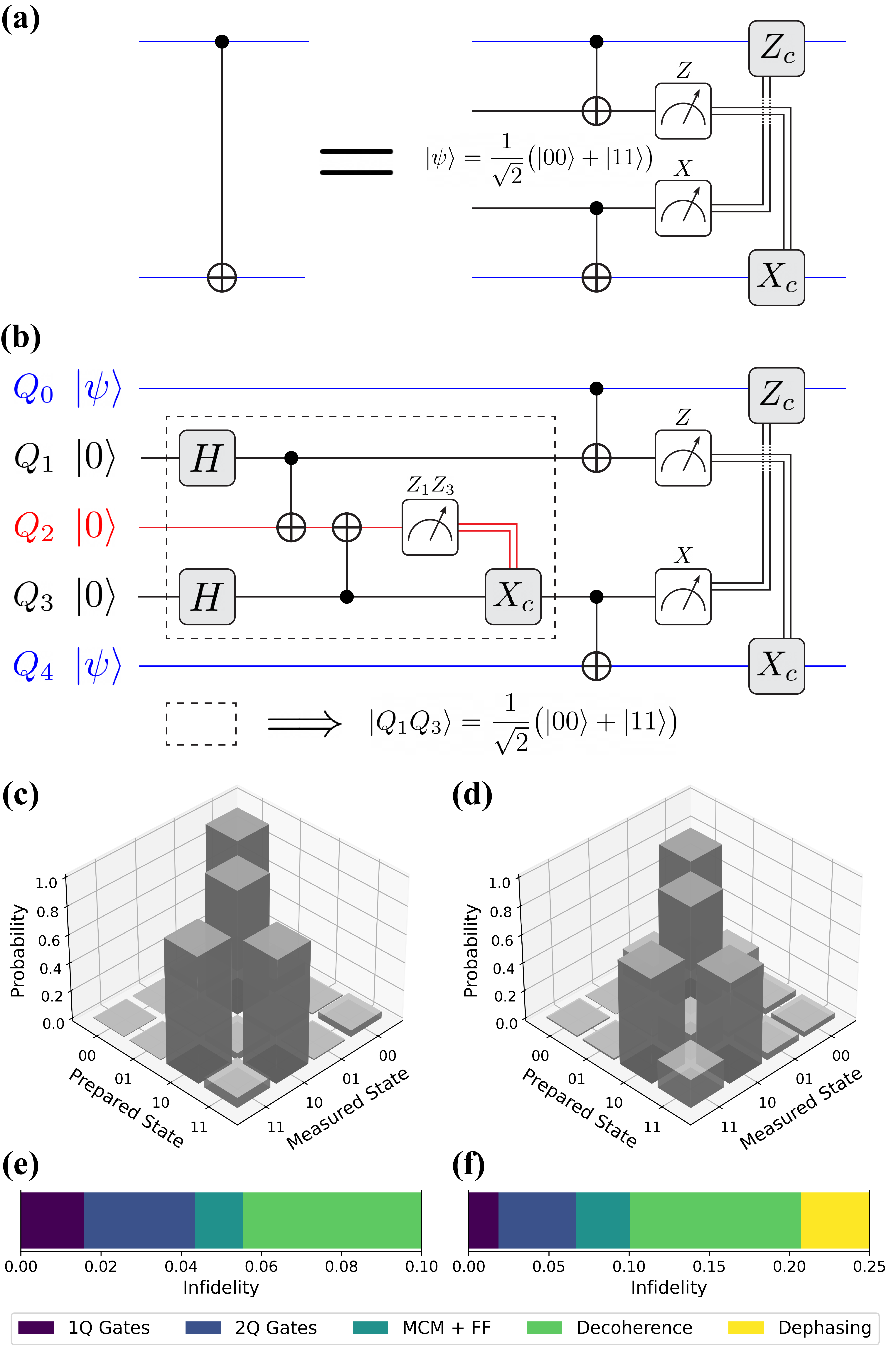}
    \caption{\textbf{Teleportation-based CNOT.} 
        \textbf{(a)} A CNOT between non-local qubits (blue) can be implemented via gate teleportation as long as the ancillae qubits 
        (black) are prepared in a Bell state.
        \textbf{(b)} When the ancillae qubits are also non-local, they can first be prepared in a Bell state using the constant depth GHZ state preparation protocol which utilizes additional ancillae (red).
        \textbf{(c)} Truth table for a teleportation-based CNOT between qubits 1 and 4. Here, the Bell state between ancillae qubits is prepared using only unitary operators. The measured fidelity is $F_\mathrm{tt} = 0.90(1)$.
        \textbf{(d)} Truth table for a teleportation-based CNOT between qubits 0 and 4. Here, the Bell state between ancillae qubits is prepared using the procedure shown in (b). The measured fidelity is $F_\mathrm{tt} = 0.75(1)$.
        \textbf{(e)} Error budget for the CNOT between qubits 1 and 4.
        \textbf{(f)} Error budget for the CNOT between qubits 0 and 4.
    }
    \label{fig:cnot}
\end{figure}

Teleportation-based quantum gates are an important component of quantum internet and networking protocols \cite{kimble2008quantum, muralidharan2016optimal, jiang2009quantum}. Moreover, such gates enable entanglement between non-local qubits on systems with sparse connectivity, which can ease topology constraints when frequency crowding is an issue \cite{morvan2022optimizing}. 
They may also be beneficial in the realm of quantum error correction. For example, for superconducting quantum processors, qubit connectivity is generally limited by locality. Therefore, for certain quantum error correcting codes that require nonlocal stabilizer measurements, the teleportation-based CNOT provides a low-depth solution for non-local entanglement\cite{delfosse_bounds_2021}.
Teleportation-based CNOT gates have been previously demonstrated in optical systems \cite{huang2004experimental}, trapped ions \cite{wan2019quantum}, superconducting cavities \cite{chou2018deterministic}, and superconducting qubits \cite{baumer2023efficient}. In our work, we demonstrate a teleportation-based CNOT gate as it serves as a precursor to the more expressive gate, the unbounded fan-out (see next section), which can also be demonstrated in constant depth using similar methods.

The circuit diagram for the teleportation-based CNOT is shown in \fig\ref{fig:cnot}(a), where it can be seen that a CNOT between distant qubits (blue) can be implemented by means of two MCMs with feed-forward, as long as the two ancillae qubits (black) are prepared in a Bell state. If the ancillae qubits themselves are non-local, then we may first prepare them in a Bell state using the same procedure introduced in the previous section for the constant-depth GHZ state preparation [\fig\ref{fig:cnot}(b)]. As a result, the ability to perform constant depth long-range GHZ preparation provides the opportunity to execute constant depth teleportation-based CNOT, regardless of the distance between two data qubits. Because we are limited by measurement-induced dephasing on spectator qubits (see Appendix \ref{sec:meas_dephasing}), we cannot characterize teleportation-based CNOT gates with high fidelity using standard methods, such as quantum process tomography. Instead, we perform truth table tomography, which is a partial tomography technique that is limited to input states prepared in the computational basis \cite{hashim2024practical}. In Figs.~\ref{fig:cnot}(c)--(d), we plot the truth table for performing a CNOT between qubits on opposite sides of our eight-qubit quantum processor, which is arranged in a ring geometry (and thus has only linear connectivity between nearest neighbors). We calculate a truth table fidelity for each pair of gates,
\begin{equation}\label{eq:tt_fid}
     F_\mathrm{tt} = \frac{1}{d} \mathrm{Tr}(S_\mathrm{exp}^{\texttt{T}} S_\mathrm{ideal}) ~,
\end{equation}
where $d = 2^n$ is the Hilbert space dimension for $n$ qubits, $S_\mathrm{exp}$ is the experimental truth table, and $S_\mathrm{ideal}$ is the ideal truth table \footnote{Equation \ref{eq:tt_fid} is quantitatively similar to the process fidelity of a CNOT gate, such as what would be estimated using quantum process tomography \cite{hashim2024practical}. In fact, a measurement of the truth table in two different bases can be used to lower bound the process fidelity \cite{hofmann2005complementary}}. For a CNOT between qubits 1 and 4 (which are only separated by two ancillae, and thus only require unitary preparation of a Bell state), we find a truth table fidelity of $F_\mathrm{tt} = 0.90(1)$. For a CNOT between qubits 0 and 4 [which are separated by three ancillae, and thus can utilize the constant depth GHZ state prep shown in \fig\ref{fig:cnot}(b)], we find a truth table fidelity of $F_\mathrm{tt} = 0.75(1)$. 

In Figs.~\ref{fig:cnot}(e) and (f), we plot the error budgets for the CNOTs between qubits 1 and 4, and 0 and 4, respectively. For the CNOT between qubits 1 and 4, we do not observe any dephasing errors --- this is due to the fact that the qubits remain in the computational basis for truth table tomography; therefore, they are insensitive to any measurement-induced dephasing. In contrast, for the CNOT between qubits 0 and 4, we observe some dephasing due to the fact that qubits 1 and 3 idle during the MCM used for the initial Bell state prep sub-circuit. In general, these fidelities and error budgets are only representative of the performance of the gate for state preparations and measurements in the computational basis; if we were able to prepare and measure in orthogonal bases, the effects of the measurement-induced dephasing would likely result in worse performance 
(see Appendix \ref{sec:meas_dephasing})
. We leave an exploration of this for future work.

Similar to the GHZ state preparation protocol, the teleportation-based CNOT is a well-known protocol \cite{bennett1993teleporting} that has previously been implemented on various different hardware platforms \cite{huang2004experimental, chou2018deterministic, wan2019quantum, baumer2023efficient}. Relevant to this work, \R\cite{baumer2023efficient} finds that the teleportation-based CNOT outperforms the equivalent unitary circuit for $\gtrsim$ 10 superconducting qubits in a linear chain. This improvement largely comes from $4\times$ reduction in the number of CNOT gates needed for the teleportation circuit compared to the unitary circuit, which requires decomposing SWAP gates into CNOTs. Moreover, they find that the performance of real-time correction is approximately equal to the equivalent measurement-based protocol with postprocessing up to about 40 qubits. This demonstrates that teleportation-based CNOTs can help alleviate topology constraints on devices with limited connectivity even on contemporary NISQ processors, effectively enabling all-to-all connectivity (as long as ancilla qubits are available). 

\section{Quantum Fan-Out Gate}\label{sec:fan-out}

\begin{figure*}[t]
    \centering
    \includegraphics[width=2\columnwidth]{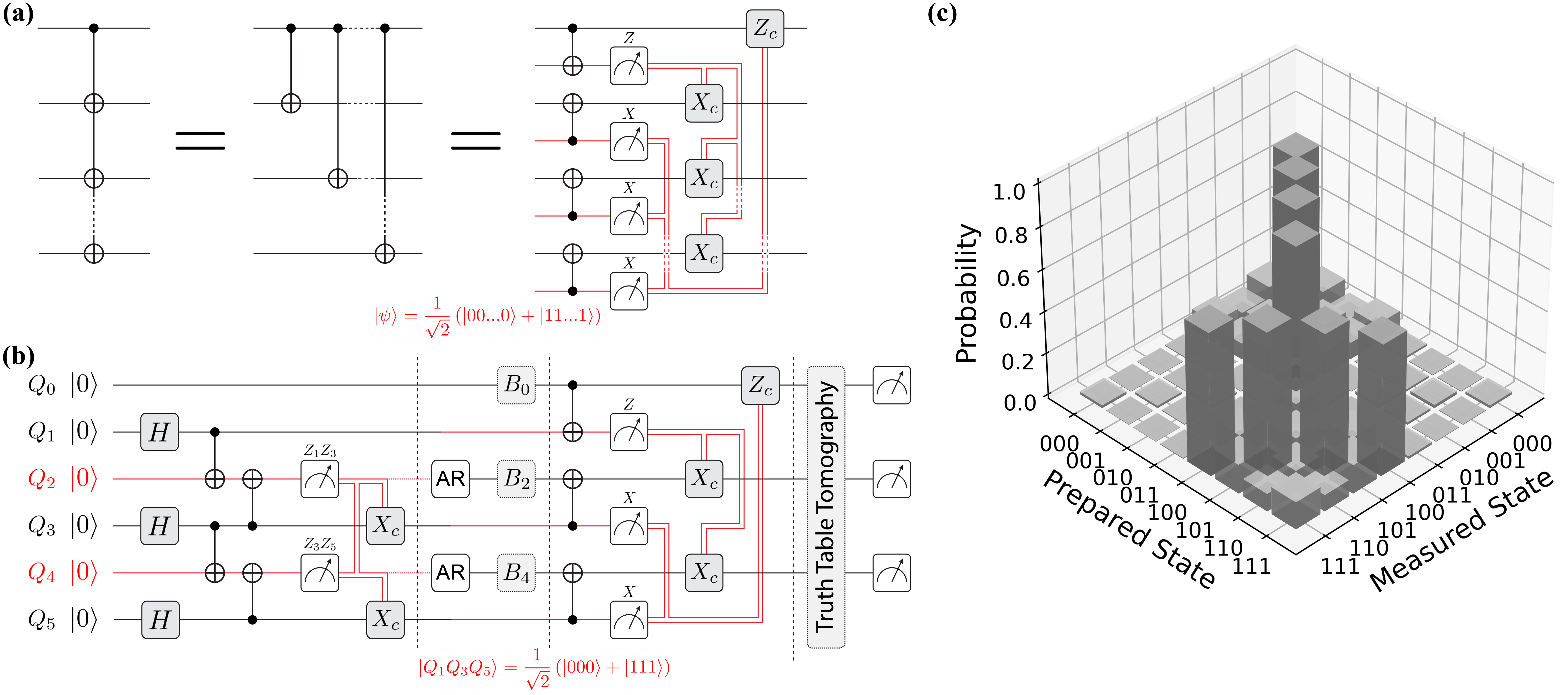}
    \caption{\textbf{Unbounded Fan-Out Gate.}
    \textbf{(a)} The quantum fan-out gate is equivalent to performing a control-NOT$^N$ on $N$ target qubits (black). Similar to the teleportation-based CNOT gate, this can be implemented in constant depth by means of MCM and feed-forward control, provided one can use $N$ ancillae qubits (red) prepared in a GHZ state as an entanglement resource.
    \textbf{(b)} We implement a controlled-NOT-NOT gate between three non-local qubits using the protocol shown in (a). We first use two ancillae qubits to prepare three other qubits in a GHZ state using the protocol shown in \fig\ref{fig:ghz}(a). Next, the ancillae qubits are actively reset (AR) as data qubits, and the GHZ qubits are used as ancillae for the fan-out gate. We perform computational basis rotations (B) on data qubits prior to the fan-out gate, and measure the data qubits in the computational basis after the fan-out gate to perform truth table tomography.
    \textbf{(c)} Experimental results for truth table tomography performed on the controlled-NOT-NOT gate. We measure a truth table fidelity of $F_\mathrm{tt} = 0.68(2)$.
    \textbf{(d)} Error budget for the controlled-NOT-NOT gate.
    }
    \label{fig:fan-out}
\end{figure*}

With slight modifications to the teleportation-based CNOT, it is possible to construct another quantum gate --- the unbounded quantum fan-out operation \cite{hoyer2005quantum} --- in constant depth with the help of MCMs and feed-forward control. The unbounded quantum fan-out [see \fig\ref{fig:fan-out}(a)] is a powerful multi-qubit gate that enables constant depth circuits to approximate various operations with a polynomially small error \cite{hoyer2005quantum}, and it attracts interest in the study of quantum circuit complexity \cite{hoyer2005quantum}. Moreover, it can help reduce circuit depth for many different tasks if it can be implemented natively, and in general gives a linear speedup over serialized instructions \cite{gokhale2021quantum}. For example, the Quantum Fourier Transformation (QFT) --- a central component of algorithms for performing discrete logarithms and factoring of integers \cite{shor1994algorithms} --- is typically performed in quadratic depth \cite{coppersmith2002approximate}, but can be approximately performed in constant depth and polynomial size using the fan-out gate \cite{hoyer2005quantum}. Other gates like $n$-qubit generalized Toffoli can also be implemented exactly in constant depth with the fan-out gate \cite{takahashi_collapse_2016}, which enables the possibility for arbitrary state preparation in constant depth together with one- and two-qubit operations (at the expense of exponential ancillae overhead) \cite{rosenthal_query_2023}. For certain states like W-state or low-Hamming-weight Dicke states, constant-depth polynomial-size preparation can be performed with MCM even when the locality constraint on two-qubit gates is considered \cite{buhrman_state_2024-1}.

The quantum fan-out gate is equivalent to performing a set of serialized controlled-NOTs with one control qubit and $N$ target qubits [see \fig\ref{fig:fan-out}(a)]. However, similar to the teleportation-based CNOT introduced in the previous section, this can be implemented in constant depth by means of MCM and feed-forward control. Our circuit for constant-depth implementation of quantum fan-out gate is shown in Fig.~\ref{fig:fan-out}(b). We first prepare a three-qubit GHZ state in constant depth as the entanglement resource (we are limited to three ancillae qubits, because at four qubits we cannot maintain phase coherence; see \fig\ref{fig:ghz}). Next, we entangle the control and all of the target qubits with the GHZ state in a single layer of CNOT gates. We then measure each ancilla qubit along $X$ or $Z$. Finally, based on the measurement results, if the $Z$ outcome is $-1$ we apply a conditional $X$ operation to all of the target qubits, and if there are an odd number of $-1$ outcomes in $X$ measurements we apply a conditional $Z$ gate on the control qubit. Due to the limited number of qubits, we first use two qubits as the ancillae in the constant-depth preparation of the GHZ state and then re-initialize them (using active reset) as the data qubits for the fan-out gate. In principle, with more qubits and better connectivity, we may choose two separate sets of ancillae qubits that each serve only one purpose. 

To demonstrate the feasibility of the fan-out gate, we implement a controlled-NOT-NOT (CXX) between three non-local qubits in constant depth. 
In \fig\ref{fig:fan-out}(c), we benchmark the CXX gate using truth table tomography and measure a truth table fidelity of $F_{tt} = 0.68(2)$. As we have previously pointed out, our qubits suffer from measurement-induced dephasing (see Appendix \ref{sec:meas_dephasing}). Performing truth table tomography to some degree is immune to this effect, since the data qubits are initialized and measured in the computational basis (after being used to prepare ancillae qubits in a GHZ state).
Consequently, we observe that the fidelity is primarily limited by decoherence and gate errors [see \fig\ref{fig:fan-out}(d)].
However, future work on other quantum processors will explore to what extent we can successfully prepare and measure in non-computational basis states (e.g., performing a controlled-Z-Z gate), as phase coherence will be important when utilizing this gate within a larger quantum computation.

After the first appearance of the current work, other works also reported implementing the fan-out gate using adaptive circuits with superconducting qubits \cite{song2025constant, baumer2025measurement}. Both works improve upon our results. Namely, \R\cite{song2025constant} implements the protocol on four qubits, and demonstrate that they are not limited by measurement-induced dephasing using complete state tomography performed on their output states when their fan-out gate is applied to both computational and superposition input states (though they note that dephasing errors and MCM errors still have a significant contribution to the error budget). They report fidelities of 0.76 -- 0.803, though we note that these are not directly comparable to our results, since theirs are \textit{state} fidelities and ours is a truth-table fidelity, which is akin to a \textit{process} fidelity. They additionally show that using real-time feed-forward incurs a larger error than simply applying Pauli-frame updates in post-processing, and that at the scale of four qubits the unitary implementation of the gate has lower error than the adaptive circuit implementation. However, \R\cite{baumer2025measurement} goes on to show that the adaptive circuit approach outperforms the unitary circuit for $n \ge 7$ qubits in a linear chain, demonstrating that such adaptive-based protocols can be beneficial even on contemporary quantum processors.
\section{Entanglement Swapping and Teleportation}\label{sec:swap}

\begin{figure*}[t]
    \centering
    \includegraphics[width=1.9\columnwidth]{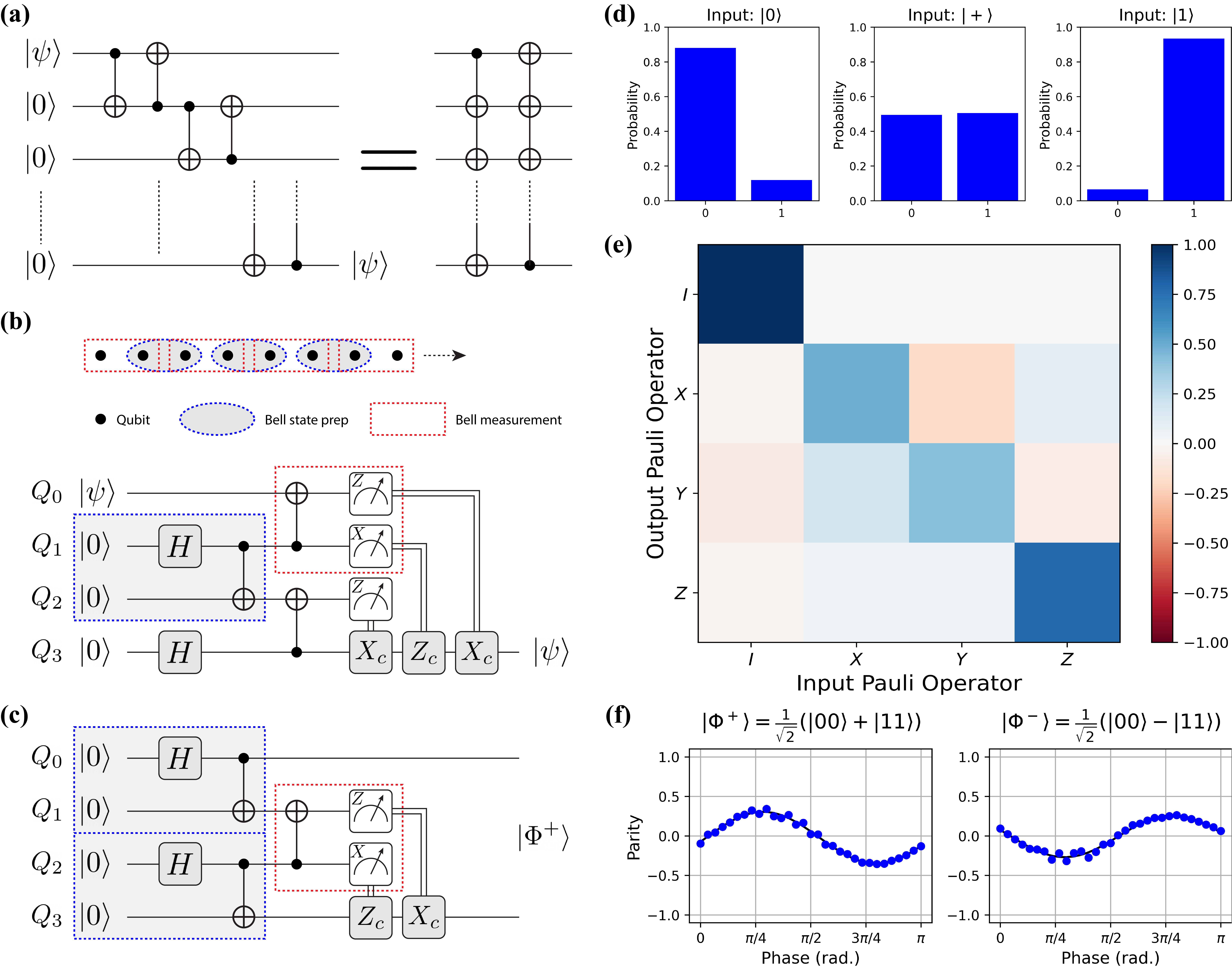}
    \caption{\textbf{Entanglement Swapping and Teleportation.} 
    \textbf{(a)} In a unitary circuit, swapping a state $\ket{\psi}$ between two ends of a register is achieved with a cascade of SWAP gates (the first CNOT gate in each SWAP can be omitted because the ancillae qubits start in $\ket{0}$). This is equivalent to two unbounded fan-out gates with the opposite orientation.
    \textbf{(b)} In measurement-based circuits, the state $\ket{\psi}$ can be \emph{teleported} from one end of a register to another using a quantum repeater protocol with simultaneous entanglement swapping, whereby a series of Bell state preparations and Bell measurements are performed in an alternating manner, and the outcome of each MCM is used to perform a conditional operation ($X_c$ or $Z_c$, if measured in the $Z$ or $X$ basis, respectively) on the final qubit.
    \textbf{(c)} A related procedure, known as entanglement swapping, can be used to deterministically prepare a Bell state between the two end qubits in a register. If the input state is $\ket{00}$, the output state is $\ket{\Phi^+}$.
    \textbf{(d)} Classical outcomes for the teleportation of $\ket{0}$, $\ket{+}$, and $\ket{1}$ from one side of our quantum processor to the other. The probabilities of successful teleportation (measured via 1 - the total variational distance to the ideal distribution) are 88.1\%, 99.4\%, and 94.4\% respectively.
    \textbf{(e)} The Pauli transfer matrix (PTM) for the teleportation protocol in (b). The ideal matrix should be $\text{diag(1, 1, 1, 1)}$, since the input and output states should be identical. The low $X$ and $Y$ eigenvalues of the PTM ($\sim0.5$) suggest there is strong (measurement-induced) dephasing.
    \textbf{(f)} Parity oscillations for the deterministic preparation of $\ket{\Phi^+}$ and $\ket{\Phi^-}$ using the protocol in (c), giving Bell state fidelities of $F_{\ket{\Phi^+}} = 0.57(1)$ and $F_{\ket{\Phi^-}} = 0.55(1)$. The small contrast in the oscillations additionally suggests a loss of phase coherence during the MCMs.
    }
    \label{fig:swap}
\end{figure*}

Similar to the teleportation-based CNOT protocol, long-distance state teleportation can also be implemented in constant depth. In \fig\ref{fig:swap}, we outline a state-teleportation protocol from the perspective of quantum repeaters \cite{jiang2009quantum, azuma2023quantum}, a useful design to tackle imperfections during long-distance entanglement distribution in quantum networks \cite{mastriani2023simplified}. In a purely unitary circuit, swapping states between two ends of a register is performed by means of a cascade of SWAP gates [or alternatively, two unbounded fan-out gates; see \fig\ref{fig:swap}(a)]. However, by utilizing MCMs and feed-forward control, it is possible to instead \emph{teleport} the state from one end of a register to another in constant depth by a quantum repeater protocol with simultaneous entanglement swapping \cite{jiang2009quantum}. The structure of quantum repeaters is simple: ancillae qubits are prepared in a Bell state in an alternating manner, and Bell measurements are performed on overlapping qubits from each Bell state [see \fig\ref{fig:swap}(b)]. The measurement of each repeater can be executed in parallel, after which conditional feed-forward operations are applied to the qubit on the opposite side of the register from the qubit whose state we wish to teleport. Throughout this process, the ancillae qubits are involved in a process known as ``entanglement swapping'' \cite{zukowski1993event, pan1998experimental, jennewein2001experimental}, since we get a long-distance entangled state from several short-distance ones. In fact, entanglement swapping generates a Bell pair distributed on two ends of the line of ancillae qubits [see \fig\ref{fig:swap}(c)], and we can then use this Bell pair for teleportation tasks and other teleportation-based operations like the CNOT we showed previously.

Historically, entanglement swapping and teleportation were performed on photonic systems \cite{schmid2009quantum, jin2015highly, tsujimoto2018high}, NMR \cite{nielsen1998complete}, and atomic systems \cite{barrett2004deterministic, riebe2008deterministic}. Deterministic teleportation has also been demonstrated on superconducting circuits \cite{steffen2013deterministic}, but previous demonstrations of entanglement swapping on superconducting qubits \cite{ning2019deterministic} were not deterministic\footnote{While \R\citenum{ning2019deterministic} claimed ``deterministic'' entanglement swapping, their protocol did not implement any feed-forward operations. Therefore, they were only able to deterministically prepare an entangled Bell state, without any active control over which state they prepared.}. Here, we demonstrate both deterministic state teleportation and deterministic remote entanglement generation via MCMs and feed-forward operations. In \fig\ref{fig:swap}(d), we plot the raw outputs for teleporting $\ket{0}$, $\ket{+}$, and $\ket{1}$ states from one side of our quantum processor to the other. The probability of measuring the incorrect state (calculated via the total variational distance to the ideal outcome) is 11.9\%, 0.6\%, and 5.6\% for  $\ket{0}$, $\ket{+}$, and $\ket{1}$, respectively. However, measuring the classical outputs of our teleportation protocol is not sufficient when the desire is to embed this sub-circuit within a larger unitary circuit, where maintaining coherence is important. For this purpose, in \fig\ref{fig:swap}(e) we plot the Pauli transfer matrix (PTM) estimated via quantum process tomography (QPT) for the state teleportation protocol, which (in contrast to the previous protocols which had multiple data qubits and, thus, a larger chance to dephase) is feasible at the level of a single qubit. Ideally, the PTM should reflect the identity operation, where the input states are prepared on a qubit on one side of the register, and the output states are measured on a qubit on the other side of the register. While we measure a process fidelity of 0.67(1) --- which is well above the classical threshold of 1/2 --- the Pauli $X$ and $Y$ eigenvalues are not well-preserved by the process. In the PTM formalism, the preservation of a given Pauli eigenvalue is directly affected by non-commuting errors \cite{hashim2024practical}. Therefore, this is further evidence that our data qubits undergo strong measurement-induced dephasing during our MCMs, even with dynamical decoupling (see Appendix \ref{sec:meas_dephasing}). Ultimately, this will limit the utility of the protocol within larger adaptive circuits.

As previously stated, when performing state teleportation, we in fact entangle the two end ancillae qubits in a Bell state via entanglement swapping. We can therefore use this procedure as an alternative to the constant-depth GHZ state introduced in Sec.~\ref{sec:ghz}, if entanglement is only desired between two non-local qubits. By varying the input state\footnote{The input to output mapping is: $\ket{00} \mapsto \ket{\Phi^+}$, $\ket{10} \mapsto \ket{\Phi^-}$, $\ket{01} \mapsto \ket{\Psi^+}$, $\ket{11} \mapsto \ket{\Psi^-}$.}, we can deterministically generate any Bell state between the two end ancillae. To demonstrate this, we deterministically prepare $\ket{\Phi^+} = \tfrac{1}{\sqrt{2}}(\ket{00} + \ket{11})$ and $\ket{\Phi^-} = \tfrac{1}{\sqrt{2}}(\ket{00} - \ket{11})$, and once again measure their coherences via parity oscillations [shown in \fig\ref{fig:swap}(f)]. We measure Bell state fidelities of $F_{\ket{\Phi^+}} = 0.57(1)$ and $F_{\ket{\Phi^-}} = 0.55(1)$. While both of these are above the required threshold of 1/2 for genuine entanglement, it is clear from the contrast in the parity oscillations that a significant amount of phase coherence is lost in the process.
\section{Conclusion and Outlook}\label{sec:outlook}

In this work, we have demonstrated a variety of protocols for generating non-local entangled states and entangling gates in constant depth. These protocols are based on teleportation- and measurement-based quantum computing, whereby one has access to an entangled resource state that can be measured in the middle of a circuit, and whose classical outcomes can be rapidly analyzed in real-time to determine which conditional feed-forward operations to apply to the unmeasured qubits. For superconducting qubits, which have relatively fast gate times and relatively short coherence times compared to other hardware platforms, the utility of such protocols depends on the ability to perform MCMs and implement real-time feedback well within the coherence times of the qubits. With access to MCMs and fast classical feedback, these protocols enable one to greatly reduce the circuit depth of many common tasks in quantum computing. This could be particularly beneficial in the NISQ era, because it enables longer circuit depths to be performed within the coherence time of the qubits. 
While many of the applications introduced in this work could have been implemented with MCMs and post-processed updates to the Pauli frame (e.g., GHZ state preparation), rather than real-time feedback, the ultimate goal of adaptive circuits is to embed them as sub-circuits (e.g., for state preparation) within larger unitary circuits that may contain non-Clifford gates; in such cases, Pauli frame updates in post-processing would not be possible. 
Moreover, such protocols have similar requirements as quantum error correction (mid-circuit measurement, real-time syndrome decoding, etc.); therefore, developing the capabilities to implement teleportation- and measurement-based protocols will be beneficial to the advancement of real-time quantum error correction.
This is particularly important in the fault-tolerant regime\cite{google_quantum_2025,caune_demonstrating_2024}, where logical non-Clifford operations are usually implemented through magic state injection. For such operations, we will need real-time error correction using the fault-tolerant measurement outcomes on the injected ancilla, from which we will apply a conditional feedforward logical Clifford (rather than a Pauli) on the state being processed, which cannot be simply postponed to the end and corrected through a Pauli frame update.

In the current era, the utility of such protocols for superconducting qubits is severely limited by dephasing on idling spectator qubits during MCMs. This dephasing can result from the static $ZZ$ coupling between neighboring qubits \cite{mitchell2021hardware} during the measurement of one qubit \cite{jurcevic2022effective}, or it can result from any unintentional direct measurement of a qubit (e.g., due to frequency crowding of the readout resonators). Depending on the strength of the dephasing, methods such as dynamical decoupling may or may not help (see Appendix \ref{sec:meas_dephasing}). 
As the design of superconducting quantum processors improves (e.g., better frequency allocation of qubits and resonators), and readout becomes both faster and higher fidelity, we anticipate that the fidelity of adaptive protocols will be equal to (or better than) the equivalent unitary circuits, making them a useful tool when circuit depth is prohibitive. 
In future work, we plan to explore ways to protect spectator qubits during MCMs. For example, in fixed-frequency systems, the static $ZZ$ coupling can be nulled by means of simultaneous off-resonant drives on both qubits \cite{mitchell2021hardware}; or, in systems with tunable couplers, this term can be turned off entirely during measurement. To combat the unintentional (direct) measurement of a spectator qubit, we plan to explore the degree to which one can ``cloak'' the qubit from the readout cavity \cite{lledo2023cloaking}. Finally, because teleportation- and measurement-based protocols are fundamentally limited by the readout fidelity of MCMs, combining them with methods of performing readout correction of MCMs \cite{hashim2023quasi} would be an interesting avenue of exploration.

\section*{Acknowledgements}\label{sec:acknowledgements}
This work was supported by the Defense Advanced Research Projects Agency under Grant Number HR0011-24-9-0358. A.H.~acknowledges financial support from the  Berkeley Initiative for Computational Transformation Fellows Program. M.Y.~and L.J.~acknowledge support from the ARO (W911NF-23-1-0077), ARO MURI (W911NF-21-1-0325), AFOSR MURI (FA9550-19-1-0399, FA9550-21-1-0209, FA9550-23-1-0338), DARPA (HR0011-24-9-0359, HR0011-24-9-0361), NSF (OMA-1936118, ERC-1941583, OMA-2137642, OSI-2326767, CCF-2312755), NTT Research, Packard Foundation (2020-71479), and the Marshall and Arlene Bennett Family Research Program.

A.H.~acknowledges fruitful discussions with Noah Goss.

\section*{Author Declarations}\label{sec:author_declarations}

\subsection*{Conflict of Interest}\label{sec:competing_interests}
All authors declare no conflict of interest.

\subsection*{Author Contributions}\label{sec:author_contributions}
A.H.~and M.Y.~devised the experiments and wrote the manuscript. A.H.~performed the experiments and analyzed the data. M.Y.~and P.G.~provided theory support. L.C.~and C.J.~fabricated the sample. N.F., Y.X., G.H., and K.N.~developed the classical control hardware used in this work. L.J.~and I.S.~supervised all work.

\section*{Data Availability}\label{sec:data_availability}
The data that supports the findings of this study are available within the article. 


\appendix

\begin{table}[t]
\renewcommand{\arraystretch}{1.5}

\centering
\resizebox{\columnwidth}{!}{
\begin{tabular}{l | r r r r r r r r }
    \hline
    \hline
    {} & Q0 & Q1 & Q2 & Q3 & Q4 & Q5 & Q6 & Q7 \\
    \hline
    $T_1$ ($\mu$s) & 96.6(2.6) & 130.0(2.7) & 142.0(3.0) & 140.0(6.3) & 77.0(5.2) & 30.4(0.95) & 55.6(1.3) & 22.5(0.32) \\
    $T_{2}^*$ ($\mu$s) & 120.0(14.0) & 41.0(7.2) & 92.0(16.0) & 61.0(6.1) & 38.0(5.4) & 8.5(1.3) & 26.0(3.7) & 39.0(1.7) \\
    $T_{2E}$ ($\mu$s) & 120.0(8.3) & 130.0(7.5) & 140.0(12.0) & 90.0(13.0) & 110.0(11.0) & 33.0(3.6) & 90.0(14.0) & 43.0(2.2) \\
    \hline
    \hline
\end{tabular}}
\caption{\textbf{Qubit Coherences.} Qubit coherence times ($T_1$, $T_{2}^*$, $T_{2E}$) are listed above.}
\label{tab:qubit_properties}

\bigskip

\centering
\resizebox{\columnwidth}{!}{
\begin{tabular}{l | r r r r r r r r }
    \hline
    \hline
    {} & Q0 & Q1 & Q2 & Q3 & Q4 & Q5 & Q6 & Q7 \\
    \hline
    $P(0 | 0)$ & 0.995(1) & 0.995(1) & 0.995(1) & 0.992(2) & 0.998(1) & 0.991(1) & 0.997(1) & 0.987(3) \\
    $P(1 | 1)$ & 0.983(2) & 0.962(7) & 0.994(2) & 0.986(2) & 0.984(5) & 0.986(2) & 0.994(2) & 0.986(2) \\
    \hline
    \hline
\end{tabular}}
\caption{\textbf{Readout Fidelities.}
Simultaneous readout fidelities for all qubits with excited state promotion.}
\label{tab:ro_fid_esp}

\bigskip

\centering
\resizebox{\columnwidth}{!}{
\begin{tabular}{l | r r r r r r r r }
    \hline
    \hline
    {} & Q0 & Q1 & Q2 & Q3 & Q4 & Q5 & Q6 & Q7 \\
    \hline
    RB iso. ($10^{-3}$) & 1.5(1) & 0.33(2) & 0.54(3) & 1.0(1) & 3.2(1) & 1.92(9) & 2.4(3) & 2.58(9) \\
    RB sim. ($10^{-3}$) & 2.1(2) & 3.1(2) & 2.0(3) & 1.9(2) & 6.1(7) & 5.7(3) & 3.4(3) & 7.6(9) \\
    \hline
    \hline
\end{tabular}}
\caption{\textbf{Single-qubit Gate Infidelities.}
    The process infidelities for isolated and simultaneous single-qubit gates measured via RB for each qubit are listed above. 
    }
\label{tab:single_qubit_gate_inf}

\bigskip

\centering
\resizebox{\columnwidth}{!}{
\begin{tabular}{l | r r r r r r r r }
    \hline
    \hline
    {} & (Q0, Q1) & (Q1, Q2) & (Q2, Q3) & (Q3, Q4) & (Q4, Q5) & (Q5, Q6) & (Q6, Q7) & (Q7, Q0) \\
    \hline
    RB iso. ($10^{-2}$) & 5.0(5) & 1.62(7) & 1.64(8) & 2.6(2) & 3.8(2) & 4.6(2) & 6.7(6) & 3.8(3) \\
    CB (CZ) ($10^{-2}$) & 1.4(1) & 0.57(1) & 0.41(1) & 0.81(4) & 1.80(8) & 2.08(3) & 2.77(8) & 1.40(7)  \\
    \hline
    \hline
\end{tabular}}
\caption{\textbf{Two-qubit Gate Infidelities.}
    The process infidelities for two-qubit RB are listed above for each qubit pair used in this work. Native (CZ) gate fidelities are measured via CB.
}
\label{tab:two_qubit_gate_inf}



\end{table}

\section{Qubit \& Readout Characterization}\label{sec:appendix}

\begin{figure}[th]
    \centering
    \includegraphics[width=0.9\columnwidth]{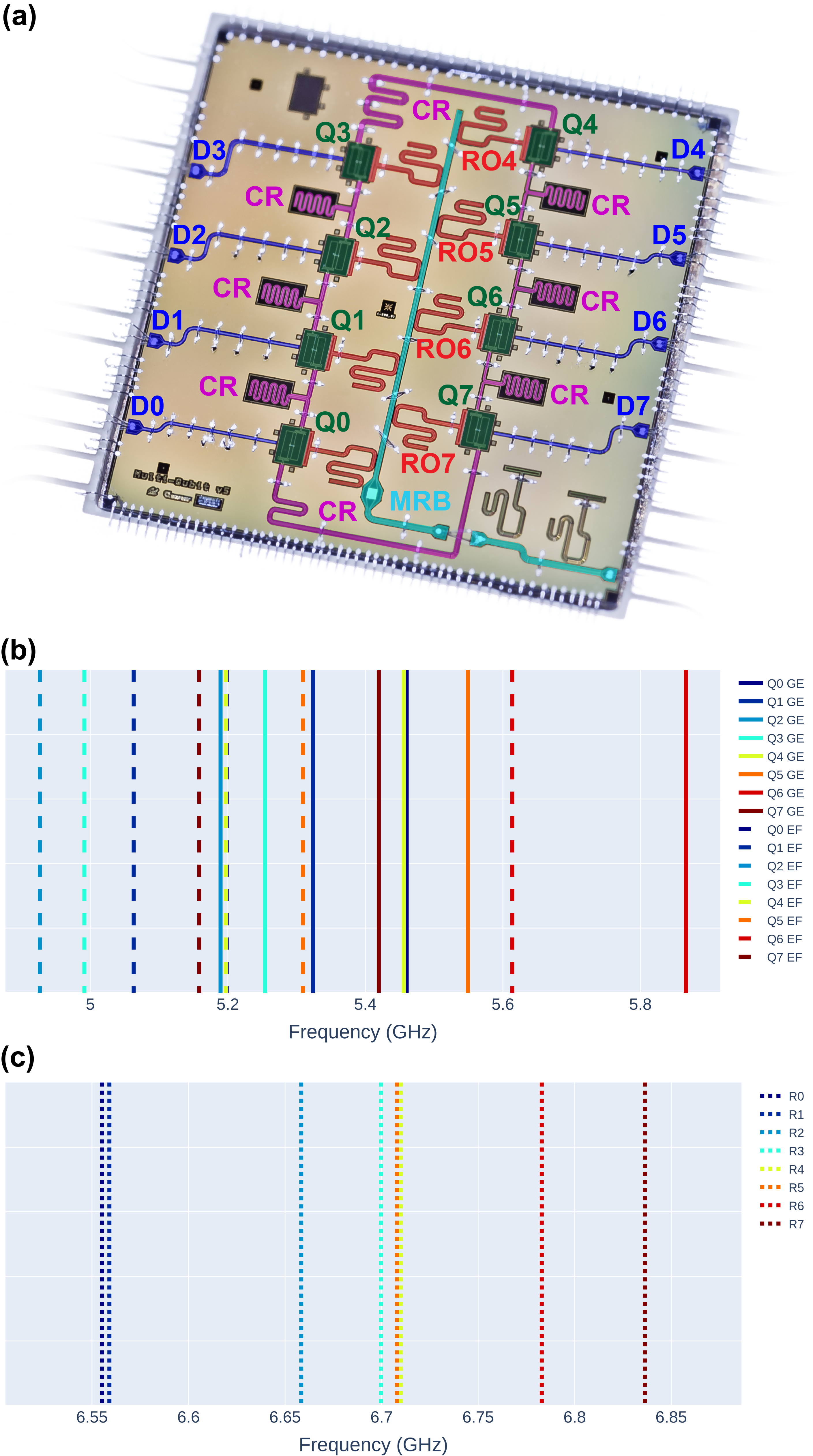}
    \caption{\textbf{Quantum Processing Unit \& Frequencies.}
        \textbf{(a)} Eight-qubit superconducting transmon processor. Qubits are labeled in green, individual drive lines are labeled in blue, individual readout resonators (RO) are labeled in red, and the multiplexed readout bus (MRB) is labeled in cyan. The qubits are coupled to nearest neighbors in a ring geometry via coupling resonators (CR, purple).
        \textbf{(b)} Qutrit frequency spectrum. The solid lines and dashed lines denote the GE and EF transition of each qutrit, respectively.
        \textbf{(b)} Readout resonator frequency spectrum. The frequency of the readout resonator coupled to each qubit is shown by a dashed line.
    }
    \label{fig:qubit_ro}
\end{figure}

The quantum processing unit (QPU) used in this work consists of eight superconducting transmon qubits arranged in a ring geometry (\fig\ref{fig:qubit_ro}a). All of the qubits on the QPU can also be operated as qutrits. The $\ket{0} \rightarrow \ket{1}$ (i.e., ``GE'') and $\ket{1} \rightarrow \ket{2}$ (i.e., ``EF'') transition frequencies of each qutrit is plotted in \fig\ref{fig:qubit_ro}b. Regions with frequency crowding can lead to microwave line crosstalk between transitions within the same subspace, or coherent leakage between transitions in different subspaces. For example, the EF transitions of Q0 and Q4 are close in frequency to the GE transition of Q2; thus, driving Q2 and result in leakage on Q0 or Q4. A similar effect can occur between the GE transition of Q5 and the EF transition of Q6, which is spectrally far from the rest of the qubits on the QPU due to fabrication inaccuracies. The GE coherence times of each qubit is listed in Table \ref{tab:qubit_properties}.

\begin{figure*}[!t]
    \centering
    \includegraphics[width=2\columnwidth]{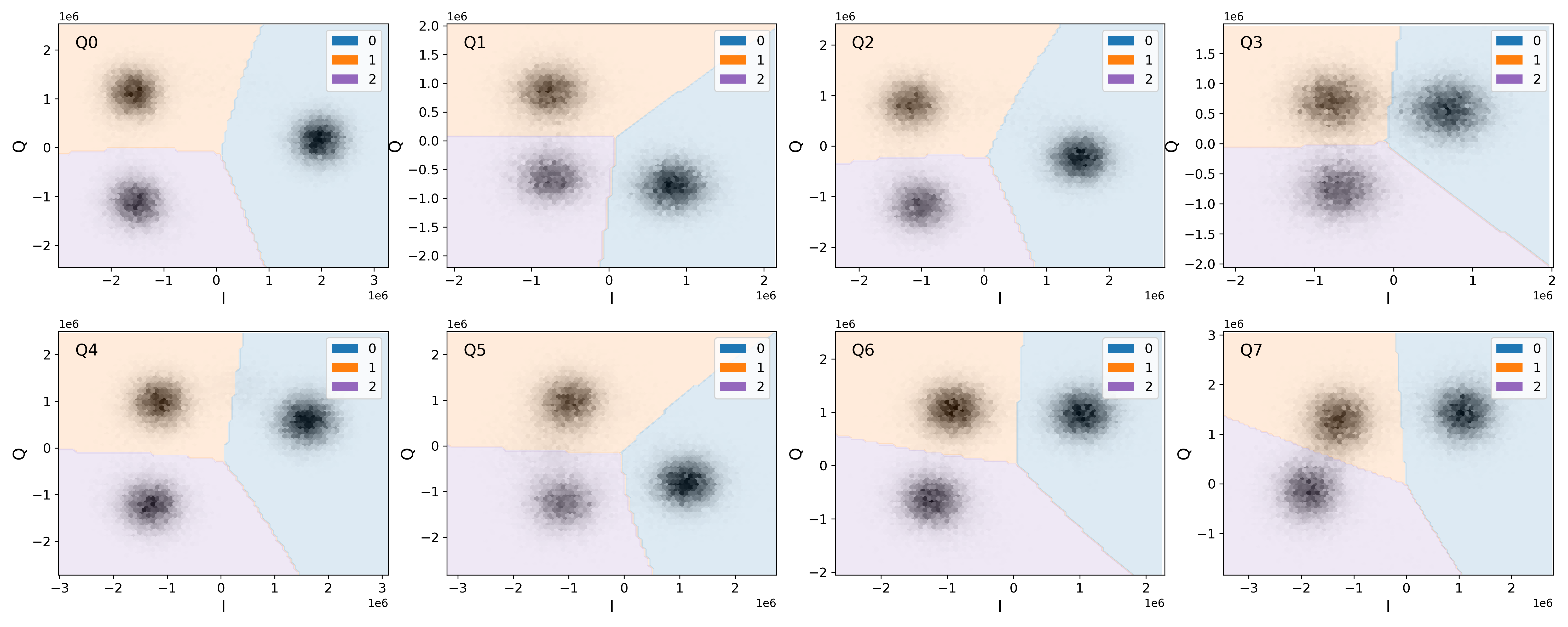}
    \caption{\textbf{Readout Calibration.} Qutrit state discrimination is supported for all qubits on the quantum processor. MCM classification is determined by the in-phase ($I$) component of the readout signal. If $I > 0$, the discriminator returns a 0; if $I < 0$, the discriminator returns a 1. Therefore, MCMs cannot currently distinguish 1 from 2, and thus cannot capture leakage.
    }
    \label{fig:ro_calibration}
\end{figure*}

The readout calibration for all eight qutrits is plotted in \fig\ref{fig:ro_calibration}, which displays the discrimination boundaries used to classify all results in this work. Qutrit state discrimination can be used to monitor leakage rates outside of the $\{\ket{0}, \ket{1}\}$ computational basis; however, this is currently limited to terminating measurements, not MCMs (see Sec.~\ref{sec:hardware} and \fig\ref{fig:ro_calibration}). Alternatively, qutrit state discrimination can be used for improving qubit readout fidelities via excited state promotion (ESP) \cite{mallet2009single}, whereby an $\pi_{1 \rightarrow 2}$ pulse is applied to each qubit before readout, after which all $\ket{2}$ state results are reclassified as $\ket{1}$ in post-processing. ESP can protect qubits against $T_1$ decay or readout-induced decay during readout. We utilize ESP to improve qubit readout both in MCMs and final measurements. The readout fidelities with ESP are listed in Table \ref{tab:ro_fid_esp}.

\section{Gate Benchmarking}\label{sec:gate_benchmarking}

The single-qubit and two-qubit gates used in this worked are benchmarked using randomized benchmarking (RB) and cycle benchmarking (CB) \cite{erhard2019characterizing}. Single-qubit Clifford infidelities measured via RB are listed in Table \ref{tab:single_qubit_gate_inf}. Two-qubit Clifford and CZ gate infidelities --- measured via RB and CB, respectively --- are listed in Table \ref{tab:two_qubit_gate_inf}. All quoted infidelities are given in terms of the process infidelity $e_F$, not the average gate infidelity $r$. These two are equal up to a simple dimensionality factor:
\begin{equation}
    e_F = \frac{d+1}{d} r \;,
\end{equation}
where $d = 2^n$ for $n$ qubits.

\section{Classical Control Hardware}\label{sec:hardware}

All experiments in this work were performed using the open-source control system \texttt{QubiC} \cite{xu2021qubic, xu2023qubic}. \texttt{QubiC} is an FPGA-based control system for superconducting qubits developed at Lawrence Berkeley National Lab. The \texttt{QubiC} system used for these experiments was implemented on the Xilinx ZCU216 RFSoC (RF system-on-chip) evaluation board, and uses custom gateware for real-time pulse sequencing and synthesis. 

The \texttt{QubiC} gateware has a bank of distributed processor cores for performing pulse sequencing, parameterization, and conditional execution (i.e., control flow) \cite{fruitwala2024distributed}. The \texttt{QubiC} readout DSP (digital signal processing) chain includes on-FPGA demodulation and qubit state discrimination using a threshold mechanism. Currently, the discrimination is performed for MCMs using the in-phase ($I$) component of the integrated readout pulse. If $I > 0$, the discriminator returns a 0; if $I < 0$, the discriminator returns a 1. For this reason, all of the $\ket{0}$ states are calibrated to be on the right side of $I = 0$, and all of the $\ket{1}$ and $\ket{2}$ states are calibrated to be on the left side of $I = 0$ (see \fig\ref{fig:ro_calibration}). These state-discriminated results can then be requested by any processor core (using a special instruction) and used as inputs to arbitrary control flow/branching decisions (e.g., a \texttt{while} loop or \texttt{if/else} code block) \cite{fruitwala2024distributed}. The total feedback latency (not including readout time) is 150 ns. After these experiments were performed, a neural network-based readout discriminator was developed for MCMs performed on \texttt{QubiC} which is capable of distinguishing $\ket{1}$ from $\ket{2}$ \cite{vora2024ml}.

\section{Look-up Tables}\label{sec:LUTs}

\begin{table}[t]
\renewcommand{\arraystretch}{1.5}

\centering
\resizebox{0.6\columnwidth}{!}{
\begin{tabular}{l | c }
    \hline
    \hline
    Outcome & Conditional Operation \\
    \hline
    0 & \\
    1 & $X_0$ or $X_2$\\
    \hline
    \hline
\end{tabular}}
\caption{\textbf{LUT for 2-qubit GHZ State.}}
\label{tab:ghz2}

\bigskip

\centering
\resizebox{0.6\columnwidth}{!}{
\begin{tabular}{l | c }
    \hline
    \hline
    Outcome & Conditional Operation \\
    \hline
    00 & \\
    01 & $X_4$ \\
    10 & $X_0$ \\
    11 & $X_2$ \\
    \hline
    \hline
\end{tabular}}
\caption{\textbf{LUT for 3-qubit GHZ State.} }
\label{tab:ghz3}

\bigskip

\centering
\resizebox{0.6\columnwidth}{!}{
\begin{tabular}{l | c }
    \hline
    \hline
    Outcome & Conditional Operation \\
    \hline
    000 & \\
    001 & $X_6$ \\
    010 & $X_0 X_2$ or $X_4 X_6$ \\
    100 & $X_0$ \\
    011 & $X_4$ \\
    101 & $X_0 X_6$ or $X_2 X_4$ \\
    110 & $X_2$ \\
    111 & $X_0 X_4$ or $X_2 X_6$ \\
    \hline
    \hline
\end{tabular}}
\caption{\textbf{LUT for 4-qubit GHZ State.} }
\label{tab:ghz4}

\end{table}

\begin{table}[t]
\renewcommand{\arraystretch}{1.5}

\centering
\resizebox{0.6\columnwidth}{!}{
\begin{tabular}{l | c }
    \hline
    \hline
    Outcome & Conditional Operation \\
    \hline
    0, 0 & \\
    0, 1 & $Z_0$ \\
    1, 0 & $X_4$ \\
    1, 1 & $Z_0 X_4$ \\
    \hline
    \hline
\end{tabular}}
\caption{\textbf{LUT for Teleportation-based CNOT.} The bit-string results for the teleportation-based CNOT do not need to be analyzed holistically; rather, they are analyzed individually for each MCM (see \fig\ref{fig:cnot}a). Thus, we do not need to nest \texttt{branch} statements on our hardware.}
\label{tab:cnot}

\bigskip

\centering
\resizebox{0.6\columnwidth}{!}{
\begin{tabular}{l | c }
    \hline
    \hline
    Outcome & Conditional Operation \\
    \hline
    0, 00 & \\
    0, 01 & $Z_0$ \\
    0, 10 & $Z_0$ \\
    0, 11 & \\
    1, 00 & $X_2 X_4$\\
    1, 01 & $Z_0 X_2 X_4$ \\
    1, 10 & $Z_0 X_2 X_4$ \\
    1, 11 & $X_2 X_4$ \\
    \hline
    \hline
\end{tabular}}
\caption{\textbf{LUT for Fan-out Gate.} The conditional $X$ gates are determined by the outcome of the first measurement, and the conditional $Z$ gate is determined by the combined outcome of the second two measurements.}
\label{tab:fan-out}

\end{table}

In this work, look-up tables (LUTs) were utilized for decision branching based on the results of the MCMs. While LUTs are not scalable as system size increases, and will eventually need to be replaced by real-time decoders, they are still viable for small system sizes. The LUTs are constructed by nesting \texttt{branch} statements on \texttt{QubiC} using the distributed processor architecture \cite{fruitwala2024distributed}. These branching statements poll the results of the measured ancillae qubits in a MCM, and then applies a conditional feed-forward operation on data qubit(s) depending on the results of the measurements. 

For example, consider the task of preparing an $n$-qubit GHZ state using the protocol introduced in Sec.~\ref{sec:ghz}. In the stabilizer formalism, a GHZ state is simultaneous eigenstate of $S_0 = X_1 X_2 ... X_n$ and $S_j = Z_j Z_{j+1}$ ($j=1,2,...,n-1$) with eigenvalue $+1$. Here, $S_j$ can be used to check the parity of any two neighboring qubits. Thus, the strategy is to initialize the data qubits in the $\ket{+}$ state so that it is an eigenstate of $S_0$ with an eigenvalue of $+1$, and then measure the stabilizers $S_j$ in parallel using a MCM on all of the ancillae qubits. The outcome of the stabilizer measurements are then analyzed holistically to determine on which data qubits we should perform a conditional bit-flip to reconstruct the $n$-qubit GHZ state. For example, suppose the data qubits are in the state $\ket{\psi} = \tfrac{1}{2}\left( \ket{000110} + \ket{111001} \right)$ prior to measurement. By entangling the data qubits pairwise with ancillae qubits, as shown in \fig\ref{fig:ghz}, we can measure the stabilizers $\{S_1, S_2, S_3, S_4, S_5\}$. In the absence of any errors, a measurement of the stabilizers will return the values $\{S_1, S_2, S_3, S_4, S_5\} = \{0, 0, 1, 0, 1\}$, where a measurement of $0$ ($1$) corresponds to a parity of $+1$ ($-1$); this tells us that we should apply a bit-flip ($X$ gate) on the 4th and 5th qubit to recover a GHZ state. Building up a LUT of all possible outcomes allows us to quickly apply conditional bit-flips in real-time, without the latency of having to decode the results on the FPGA or on some other (potentially non-local) classical processor. Though, it is worth noting that a real-time decoder of parity measurements could be implemented by analyzing the parity measurements sequentially (per clock cycle) and considering each 1 measurement as a domain wall which takes us from \textit to \textit{apply $X$}, and back again.

In Tables \ref{tab:ghz2} -- \ref{tab:fan-out}, we include the LUTs used in this work for the various experiments performed in the main text. For each experiment, we list the various possible outcomes of the parity measurements, and for each outcome we list the possible conditional operations which will recover the intended target state (preference is given to the conditional operations with the minimal number of gates). Note that all data qubits are indexed from 0 in these tables. Note also that the bit-strings are read left-to-right in accordance with increasing ancilla qubit number label. We omit a LUT for the entanglement swapping and teleportation protocols (Sec.~\ref{sec:swap}), since, like the teleportation-based CNOT, we do nothing (apply the conditional gate operation) if we measure (0) 1.

\section{Measurement-Induced Dephasing}\label{sec:meas_dephasing}

\begin{figure}[ht]
    \centering
    \includegraphics[width=\columnwidth]{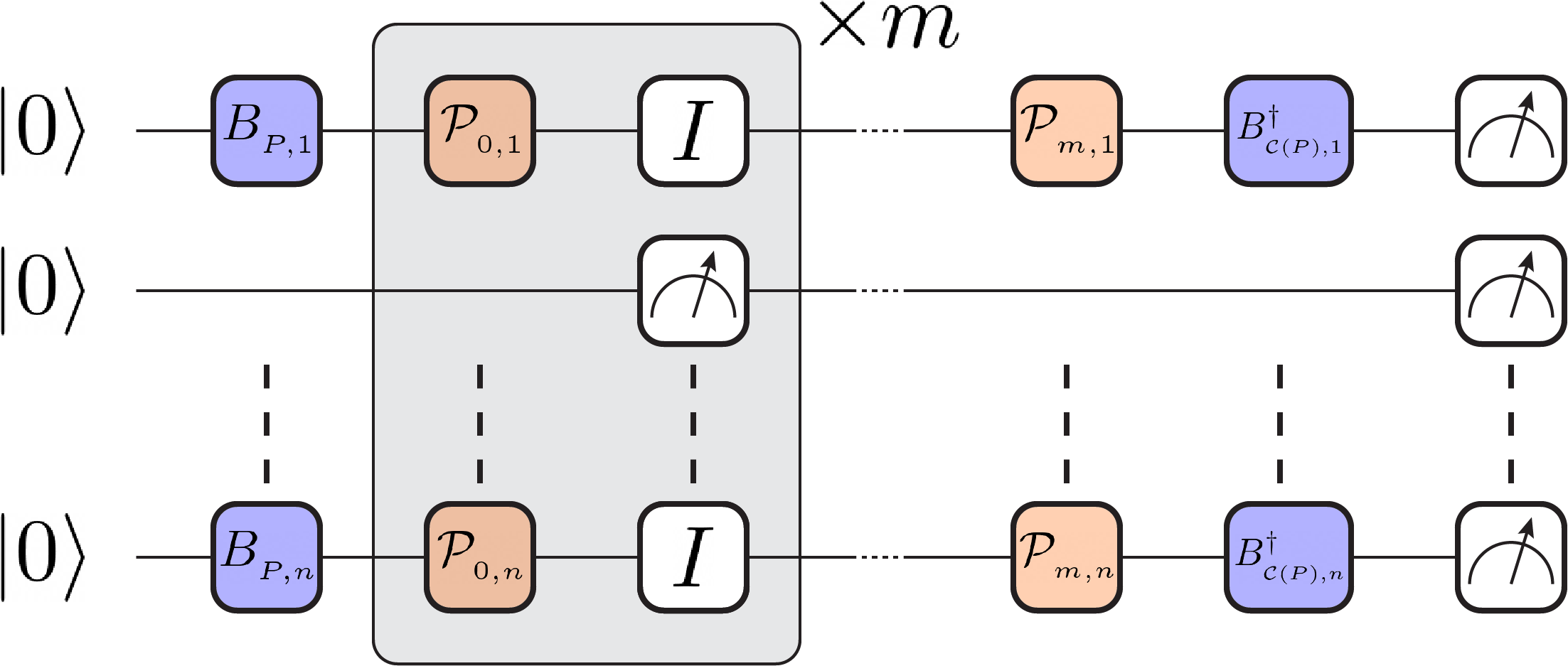}
    \caption{\textbf{Benchmarking the Impact of Mid-Circuit Measurements.} To measure the impact of MCMs on idling spectator qubits, we perform CB on the spectator qubits with an interleaved layer consisting of a MCM on another qubit. Here, $B_P$ is rotation to +1 eigenstate of the basis of the Pauli $P$, $\mathcal{P}$ is a randomly sampled Pauli gate, $m$ is the number of interleaved cycles, and $B^\dagger_{C(P)}$ rotates the system back to the original basis state.
    }
    \label{fig:cb_mcm}
\end{figure}

Mid-circuit measurements (MCMs) can lead to unwanted dephasing on nearby idling spectator qubits. This can happen in two different ways: firstly, when an ancilla qubits is measured, the static $ZZ$ coupling \cite{mitchell2021hardware} between coupled qubits can dephase the spectator qubit during the measurement process \cite{jurcevic2022effective} if the ancilla qubit undergoes a measurement-induced state transition \cite{dumas2024measurement}. This only results in dephasing between qubits that are directly coupled, and assumes that the $ZZ$ coupling cannot be turned off during the MCM. This is the case for the qubits used in this experiment, as they are coupled via fix-frequency resonators in a ring geometry (see \fig\ref{fig:qubit_ro}a). However in systems with tunable couplers, this $ZZ$ coupling can be largely suppressed, which would help alleviate the dephasing seen by the spectator qubit.

The other way in which MCMs can dephase spectator qubits is via unintentional direct measurement, i.e., measurement-induced dephasing \cite{gambetta2006qubit}. When the readout resonator frequencies of nearby qubits are not spectrally well-resolved, the measurement of one qubit can inadvertently result in the measurement of other qubits. This is the case for multiple pairs of qubits on the current device (see \fig\ref{fig:qubit_ro}c). For example, the readout resonators coupled to Q0 and Q1 are within $\sim$4 MHz of each other, and the readout resonators coupled to Q3, Q4, and Q5 are within $\sim$11 MHz of each other. This is not done intentionally; rather, it is the result of inaccuracies in the fabrication process. This directly limits which qubits can be used as ancillas for MCMs and which can be used as data qubits, because the measurement of the ancilla qubit can result in a complete loss of phase coherence on data qubits. 

\begin{figure*}[!t]
    \centering
    \includegraphics[width=1.7\columnwidth]{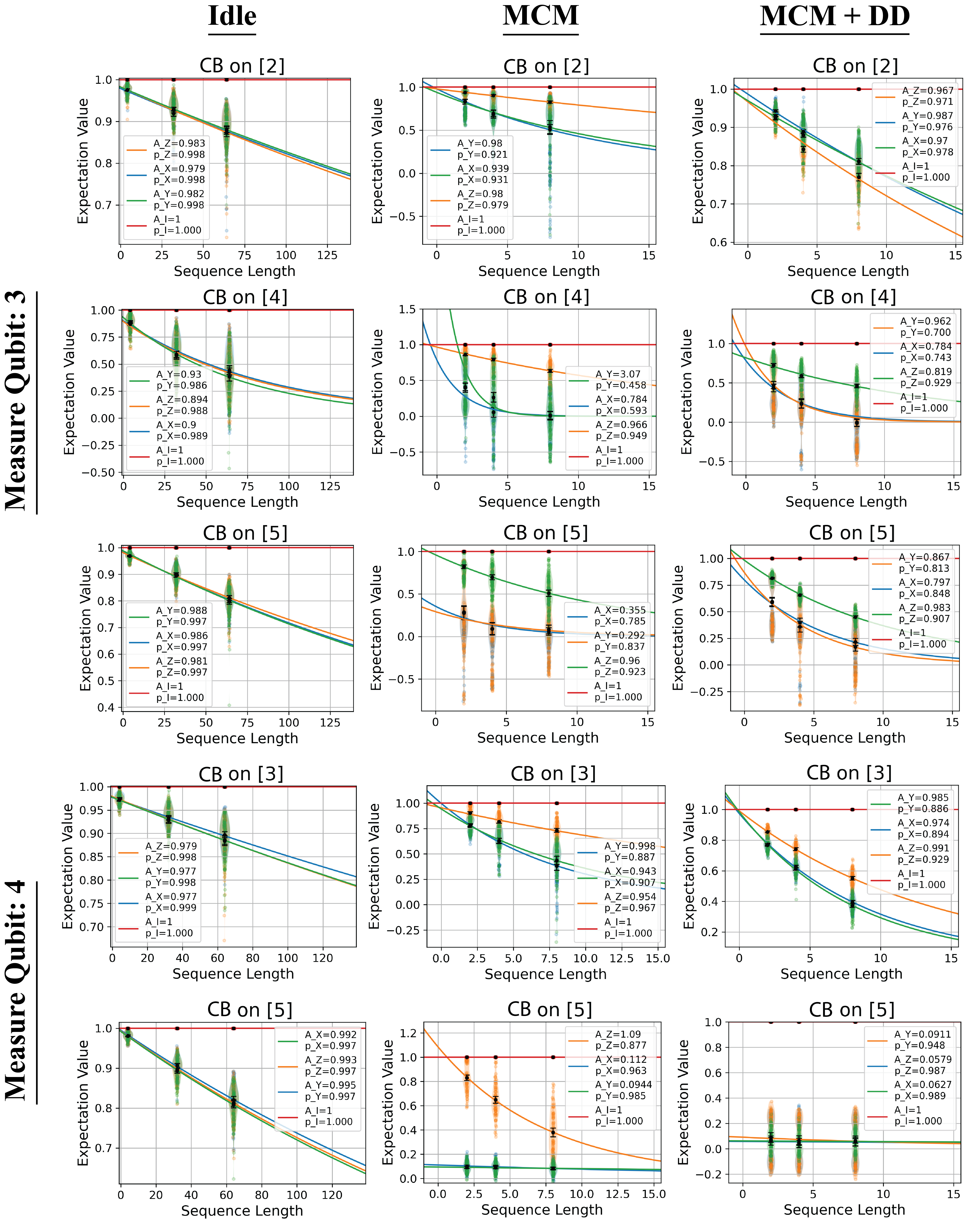}
    \caption{\textbf{Measurement-Induced Dephasing.} We observe that the measurement of Q3 has a small effect on Q2, but almost completely dephases Q4 and Q5 (note that the Pauli-$X$ and Pauli-$Y$ decays for Q4 could not even be properly fit). This is not surprising, because the readout frequencies of Q3, Q4, and Q5, are only separated by $\sim$11 MHz (see \fig\ref{fig:qubit_ro}c). While DD does help with the dephasing on Q2, it has minimal impact on Q4 and Q5. Similarly, we observe that the measurement of Q4 strongly affects Q3, and completely dephases Q5. DD helps Q3, but completely decoheres Q5.
    }
    \label{fig:dephasing2}
\end{figure*}

To measure the impact of MCMs on neighboring spectator qubits, we utilize cycle benchmarking (CB) with a slight variation (see \fig\ref{fig:cb_mcm}). Here, the interleaved cycle is a MCM on one qubit, with all of the other qubits left idling. However, because the goal is to quantify the impact of MCMs on spectator qubits, we only twirl the spectator qubits, we do not twirl the qubit being measured. A similar strategy was done in \R\citenum{govia2023randomized} using RB, but RB cannot distinguish the impact of errors along different measurement axes. The utility of CB is that it enables one to quantify how a MCM affects other qubits along $X$, $Y$, and $Z$. Moreover, we can glean information from both the state-preparation and measurement (SPAM) constant $A_P$ for the exponential decay of each Pauli operator $P$ (termed ``Pauli decays''), as well as from the exponential fit parameter $p_P$. If the effect of a MCM on an idle qubit is small, then one should be able to accurately fit $A_P$ and $p_P$. However, if the effect of a MCM is large enough such that no useable information can be gleaned for a given $P$, then $A_P$ will be small (close to zero) and $p_P$ cannot be trusted (assuming an exponential can even be fit to the data, which is not always the case). This is the case if the MCM completely dephases the idle qubit, for which $A_X \approx 0$ and $A_Y \approx 0$ (see, e.g., the bottom row of \fig\ref{fig:dephasing2}). When an exponential can be fit to the data, $p_P$ gives the probability with which $P$ is preserved by the error process. More generally, any Pauli error $Q$ will attenuate the probability $p_P$ for any non-commuting Pauli operator $P$ \cite{hashim2024practical}. Thus, if a MCM causes a spectator qubit to dephase (which is a $Z$-type error), this will affect $p_X$ and $p_Y$ (but will not affect $p_Z$). Therefore, we often observe that $p_Z > p_X, p_Y$, which is a strong indication of measurement-induced dephasing.

\begin{figure*}[!t]
    \centering
    \includegraphics[width=1.9\columnwidth]{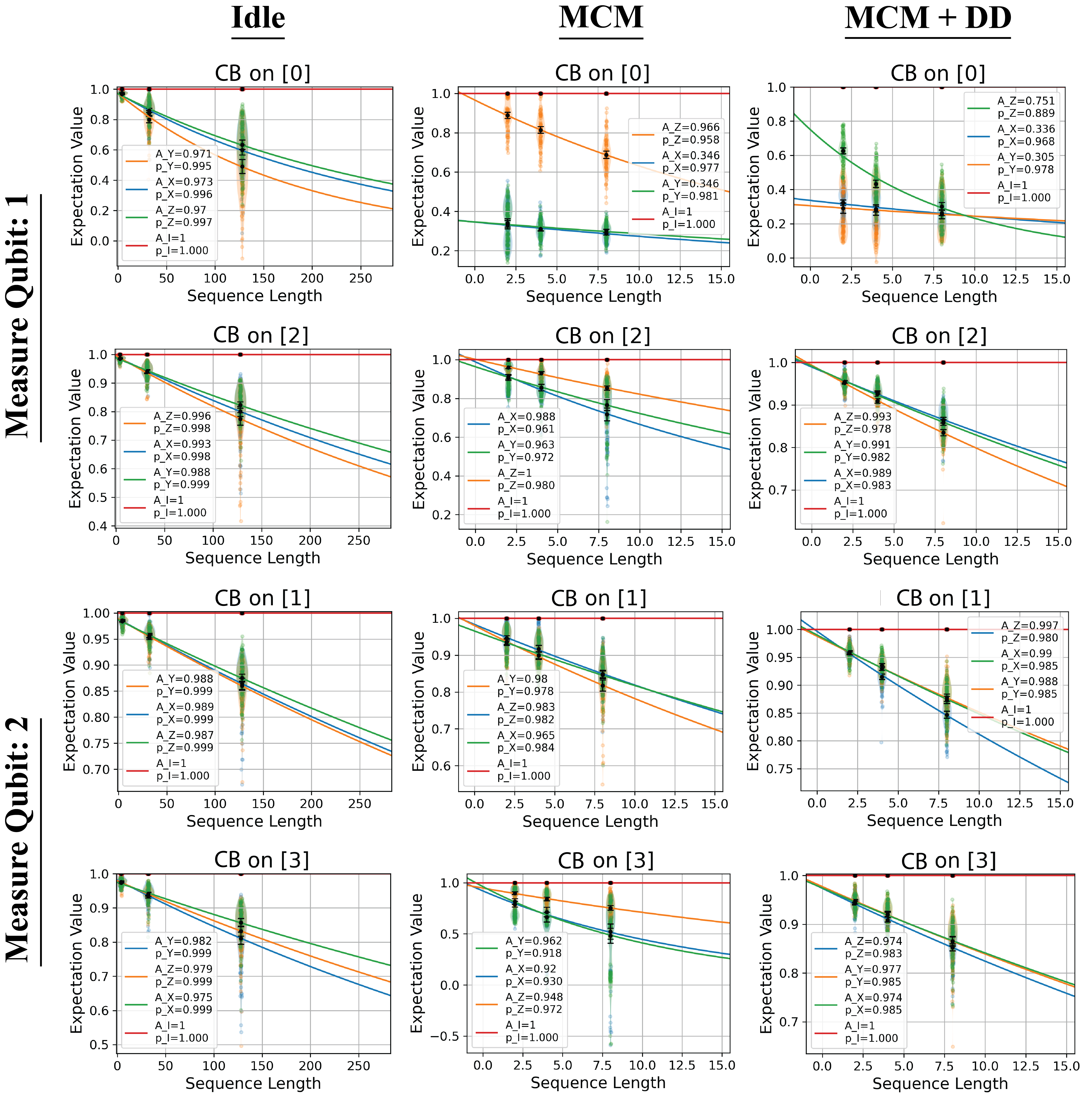}
    \caption{\textbf{Measurement-Induced Dephasing.} We observe that the measurement of Q1 almost completely dephases Q0; this is expected, because the readout resonator frequencies of Q0 and Q1 are only separated by $\sim$4 MHz (see \fig\ref{fig:qubit_ro}c). In this case, DD applied to Q0 has an adverse effect on the Pauli-$Z$ decay. However, the measurement of Q1 has a much smaller effect on Q2, and DD applied to Q2 helps mitigate the dephasing. In contrast, the measurement of Q2 has a small effect on both Q1 and Q3. DD applied to both qubits helps mitigate the dephasing.
    }
    \label{fig:dephasing1}
\end{figure*}

\begin{figure*}[!t]
    \centering
    \includegraphics[width=1.9\columnwidth]{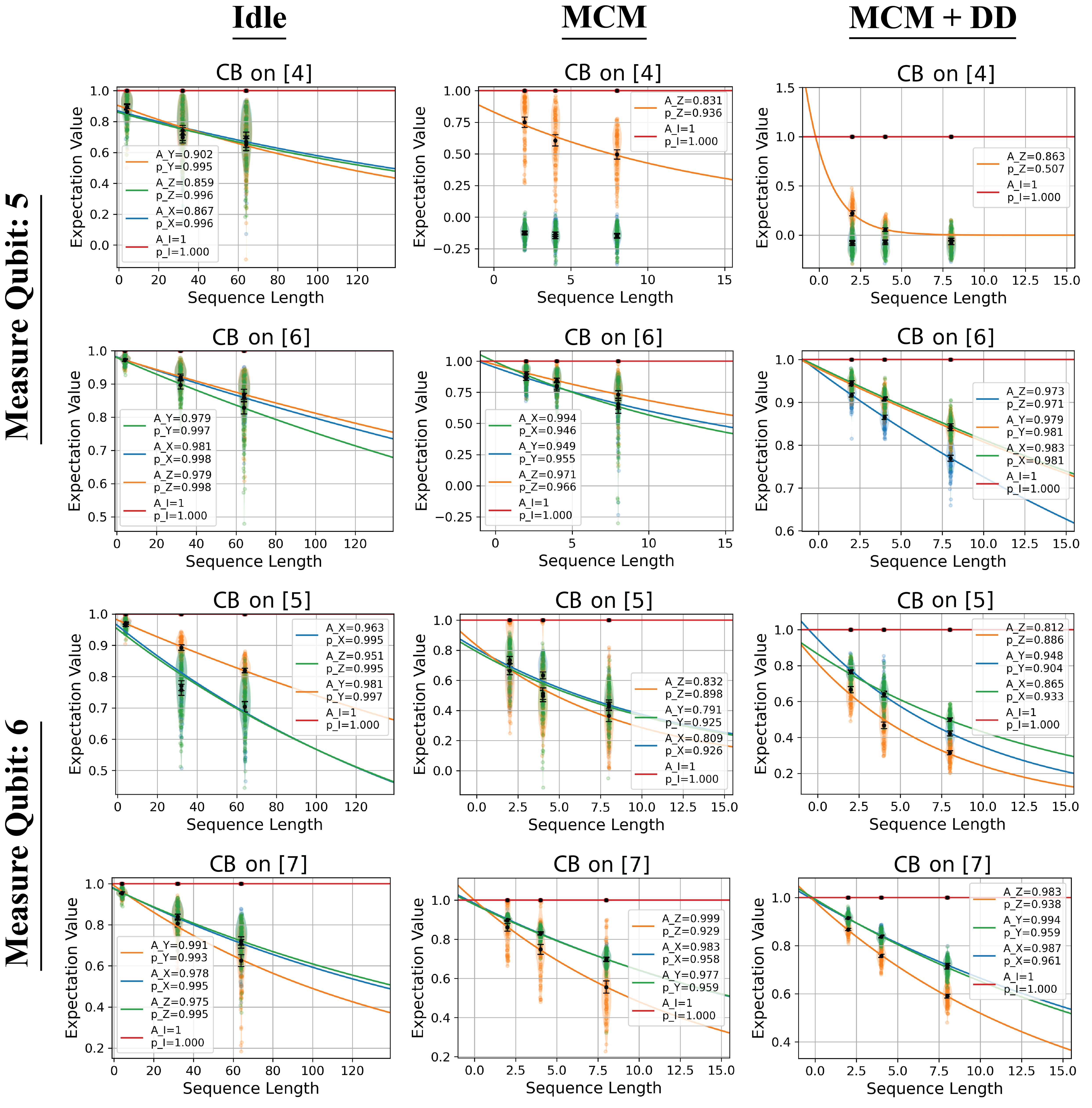}
    \caption{\textbf{Measurement-Induced Dephasing.} Similar to what we observed in \fig\ref{fig:dephasing2}, the measurement of Q5 completely dephases Q4, and DD has an adverse effect. However, Q6 is only marginally affected by Q5, and DD does indeed help. However, on the contrary, the measurement of Q6 seems to mildly dephase Q5, and DD does not appear to help. This is surprising, because the qubit and readout frequencies spectrally well-resolved. The measurement of Q6 has a smaller effect on Q7, and DD only marginally helps.
    }
    \label{fig:dephasing3}
\end{figure*}

\begin{figure*}[!t]
    \centering
    \includegraphics[width=1.9\columnwidth]{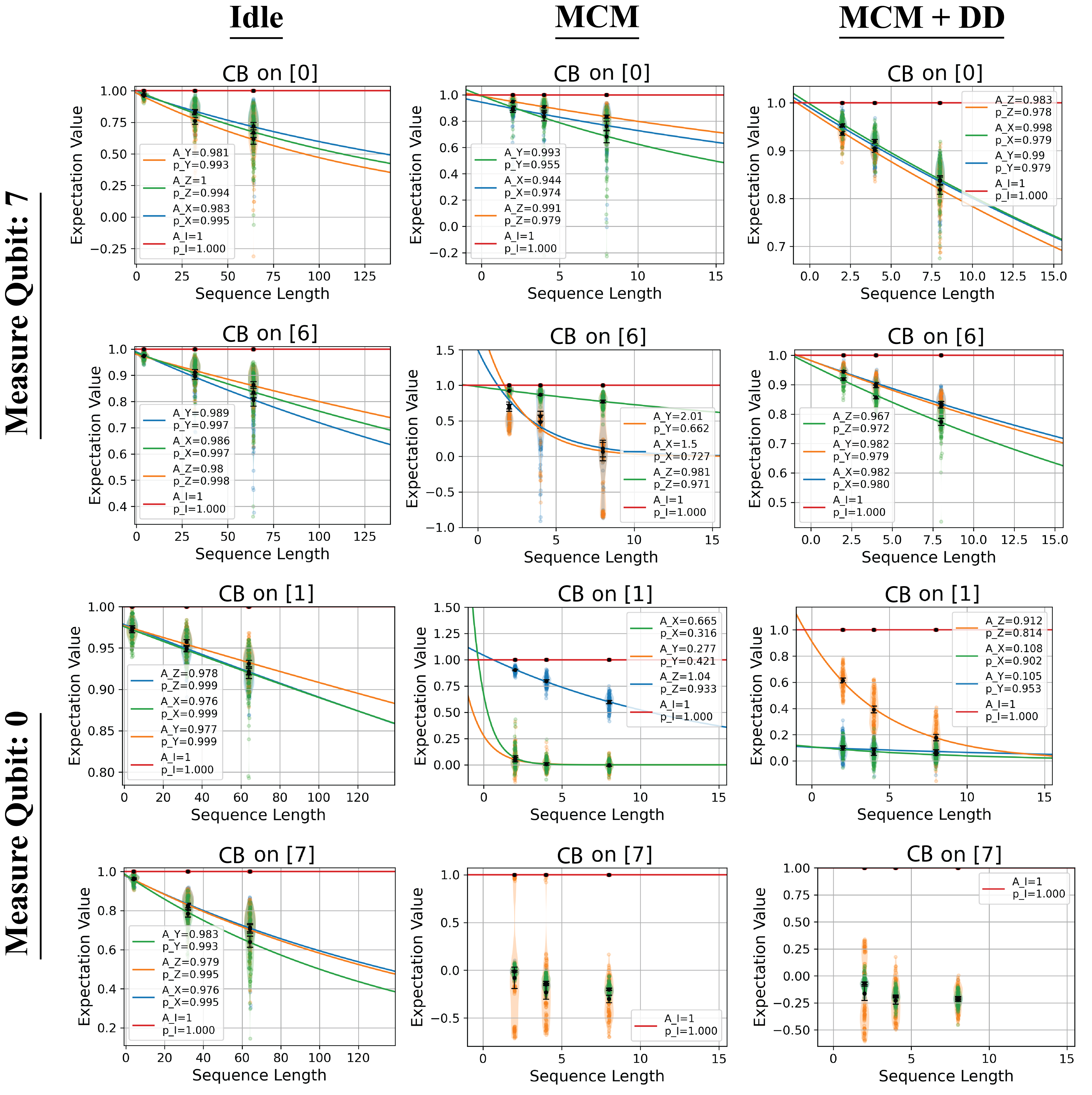}
    \caption{\textbf{Measurement-Induced Dephasing.} We observe that the measurement of Q7 has a small effect on Q0 and that DD helps mitigate the dephasing. Q7 has a much larger effect on Q6, but DD mitigates most of the effect. In contrast, the measurement of Q0 completely dephases Q1 and completely decoheres Q7. DD has an adverse effect on Q1 and no effect on Q7.
    }
    \label{fig:dephasing4}
\end{figure*}

Using CB, we systematically characterize the effect of MCMs on all nearest-neighbor idling qubits across our full device (see Figs.~\ref{fig:dephasing1}--\ref{fig:dephasing4}). We compare the effect of MCMs to the case when the spectator qubits idle (with no MCM) for the duration of the measurement, as well as the case when dynamically decoupling (DD; specifically, an XY8 sequence \cite{maudsley1986modified, ahmed2013robustness}) is applied to the spectator qubits during the MCM. In all cases, we observe that the MCM has at least \emph{some} effect on the spectator qubits. If the effect is small, we observe that DD helps reduce the dephasing (i.e., increase $p_X$ and $p_Y$). However, if the effect is large, DD does not help and can even impact the Pauli $Z$ decay adversely (see, e.g., the bottom row of \fig\ref{fig:dephasing2}). This is because DD relies on the assumption that the dephasing noise is quasi-static (i.e., slowly varying) over the course of an experiment, such that one can ``echo away'' the phase error. However, when a MCM is strong enough to inadvertently measure a nearby qubit, this breaks the underlying assumption of DD, as measurement back-action introduces strong stochastic phase rotations \cite{kockum2012undoing}. In such cases, a DD sequence like XY8 will not return to the qubit to the original state, inevitably introducing $X$- and $Y$-type errors (and thus affecting the Pauli-$Z$ decays).

\bibliography{bibliography}

\end{document}